\documentstyle[12pt,a41,eucal,epsfig,cite,amsmath,afterpage,draft]{article}
\newtheorem{example}{Example}

\newcommand{\SigmaP}{\textsf{Sigma}}
\newcommand{\mathP}{\textsf{Mathematica}}
\newcommand{\FORMP}{\textsf{FORM}}
\newcommand{\SE}{{\mathbf S}}

\newcommand{\bea}{\begin{eqnarray}}
\newcommand{\bq}{\begin{equation}}
\newcommand{\eea}{\end{eqnarray}}
\newcommand{\eq}{\end{equation}}

\newcommand{\lsim}{\raisebox{-0.07cm   }
{$\, \stackrel{<}{{\scriptstyle\sim}}\, $}}

\newcommand\be{\begin{eqnarray}}
\newcommand\ee{\end{eqnarray}}

\newcommand\Mvec{\,\mbox{\bf M}}
\begin{document}
\noindent
\sloppy
\thispagestyle{empty}
\begin{flushleft}
DESY 09-002
\hfill {\tt arXiv:0902.4091 [hep-ph]}
\\
SFB/CPP-09-22\\
February 2009
\end{flushleft}
%
%\setcounter{page}{0}
% 1
%\mbox{}
\vspace*{\fill}
\hspace{-3mm}
{\begin{center}
{\Large\bfseries Determining the closed forms of the \boldmath $O(a_s^3)$ 
anomalous}

\vspace*{2mm}
{\Large\bfseries dimensions and Wilson coefficients from}

\vspace*{3mm}
{\Large\bfseries Mellin moments by means of computer algebra}

\end{center}
}

\begin{center}
\vspace{2cm}
\large
Johannes Bl\"umlein$^a$, Manuel Kauers$^b$, Sebastian Klein$^a$, and Carsten
Schneider$^b$
\\
\vspace{5mm}
\normalsize
{\itshape $^a$~Deutsches Elektronen--Synchrotron, DESY,\\
Platanenallee 6, D--15738 Zeuthen, Germany}
\\

\vspace{5mm}
\normalsize {\itshape $^b$~Research
Institute for Symbolic Computation (RISC),\\ Johannes Kepler
University, Altenbergerstra\ss{}e 69, A-4040 Linz, Austria} \\ \vspace{2em}
%\today
\end{center}

\vspace*{\fill} %
%%%%%%%%%%%%%%%%%%%%%%%%%%%%%%%%%%%%%%%%%%%%%%%%%%%%%%%%%%%%%%%%%%%%%%%%
\begin{abstract}
\noindent
Single scale quantities, as anomalous dimensions and hard scattering cross
sections, in renormalizable Quantum Field Theories are found to obey difference
equations of finite order in Mellin space. It is often easier to calculate
fixed moments for these quantities compared to a direct attempt to derive them
in terms of harmonic sums and their generalizations involving the Mellin
parameter $N$. Starting from a sufficiently large number of given moments, we
establish linear recurrence relations of lowest possible order with polynomial
coefficients of usually high degree. Then these recurrence equations are solved in
terms of d'Alembertian solutions where the involved nested sums are represented
in optimal nested depth. Given this representation, it is then an easy task to
express the result in terms of harmonic sums. In this process we compactify the
result such that no algebraic relations occur among the sums involved. We demonstrate
the method for the QCD unpolarized anomalous dimensions and massless Wilson
coefficients to 3--loop order treating the contributions for individual color
coefficients. For the most complicated subproblem 5114 moments were needed in order to
produce a recurrence of order 35 whose coefficients have degrees up to~938. About
four months of CPU time were needed to  establish and solve the recurrences for the
anomalous dimensions and Wilson coefficients on a 2~GHz machine requiring less than
10~GB of memory. No algorithm is known yet to provide such a high number of moments for
3--loop quantities. Yet the method presented shows that it is possible to establish and
solve recurrences of rather large order and degree, occurring in physics problems,
uniquely, fast and reliably with computer algebra.
\end{abstract}
%%%%%%%%%%%%%%%%%%%%%%%%%%%%%%%%%%%%%%%%%%%%%%%%%%%%%%%%%%%%%%%%%%%%%%%%
\vspace*{\fill} \newpage
%%%%%%%%%%%%%%%%%%%%%%%%%%%%%%%%%%%%%%%%%%%%%%%%%%%%%%%%%%%%%%%%%%%%%%%%
%%%%%%%%%%%%%%%%%%%%%%%%%%%%%%%%%%%%%%%%%%%%%%%%%%%%%%%%%%%%%%%%%%%%%%% %
%           Introduction
%%%%%%%%%%%%%%%%%%%%%%%%%%%%%%%%%%%%%%%%%%%%%%%%%%%%%%%%%%%%%%%%%%%%%%%
\section{Introduction}
\label{sec:1}
%%%%%%%%%%%%%%%%%%%%%%%%%%%%%%%%%%%%%%%%%%%%%%%%%%%%%%%%%%%%%%%%%%%%%%%

\vspace{1mm}
\noindent
Precision predictions for observables in Elementary Particle Physics require
the calculation of the corresponding Feynman diagrams, the number of which grows
fast with the order in the coupling constant being considered. According to the
relevant number of different ratios of Lorentz invariants or scales involved one may
group these observables into 0-scale, 1-scale, 2-scale etc processes. In
renormalizable Quantum Field Theories the radiative corrections to the couplings,
masses and external fields are examples for 0-scale quantities \cite{X1}.
Anomalous dimensions and hard scattering cross sections, as the Wilson
coefficients for light and heavy flavors (for $Q^2 \gg m_H^2$) in deeply inelastic
scattering, are single scale quantities, cf. \cite{X2,MVV1,MVV2,MVV3,HEAV}. Also
the sub-system cross sections for the Drell-Yan process and the cross section for
hadronic Higgs-boson production in the heavy mass limit for the top--quark 
belong to this class. Mellin
moments for single scale quantities $f(x)$,
%-----------------------------------------------------------------------------
\begin{eqnarray}\label{eq:1}
\Mvec[f(x)](N) = \int_0^1~dx~x^{N}~f(x)
\end{eqnarray}
%-----------------------------------------------------------------------------
are 0-scale quantities again for $N\in {\bf N}$
\cite{LRV,MOM,MOM1,HMOM}. Here $x$
usually denotes a fraction of Lorentz-invariants the support of which is
or can be extended to $[0,1]$. In the lower order in perturbation theory 0-scale
quantities  can be expressed as linear combinations of specific numbers over
${\bf Q}$ which are multiple $\zeta$-values \cite{MZV},
%----------------------------------------------------------------------------
\begin{equation*}
\zeta_{a_1, \dots, a_n} = \sum_{k=1}^{\infty} \frac{{\rm
sign}(a_1)^k}{k^{|a_1|}} S_{a_2, \dots, a_n}(k),\quad a_i\in{\bf Z} \setminus \{0\}
\end{equation*}
%----------------------------------------------------------------------------
at the beginning, with possible extensions in higher orders, which
occur in both massive and massless calculations~\cite{X4}.
The 1-scale quantities can be expressed in terms of finite harmonic sums
\cite{HSUM,summer}
%----------------------------------------------------------------------------
\begin{equation*}
S_{a_1, \dots, a_n}(N) = \sum_{k=1}^{N} \frac{{\rm
sign}(a_1)^k}{k^{|a_1|}} S_{a_2, \dots, a_n}(k),\quad S_{\emptyset} = 1,\quad a_i\in{\bf
Z} \setminus
\{0\}
\end{equation*}
%----------------------------------------------------------------------------
and rational functions of the Mellin variable $N$ at lower orders in
perturbation theory. At higher orders one expects to find generalizations
of harmonic sums. Much less is known on the function-spaces spanning 2- and
higher scale processes. The Mellin-transformation \eqref{eq:1} is empirically found
to yield considerable structural simplifications of 1-scale processes,
cf.~\cite{COHS}. In massless processes this is partly due to the factorization
properties, but it seems to hold to an even wider extent. Corresponding
diagonalizations for processes with a higher number of scales depend on their
respective main symmetries, which may not even be fixed by just the number of
scales.

In the present paper we study single scale processes and represent them in
Mellin space. In order to apply our method under consideration, we shall assume
the case that $\Mvec[f(x)](N)$ can be found as the solution of a linear recurrence
equation
%----------------------------------------------------------------------------
\begin{eqnarray}\label{eqDI}
a_0(N)F(N)+a_1(N)F(N+1)+\cdots+a_l(N)F(N+l)=0,
\end{eqnarray}
%----------------------------------------------------------------------------
with polynomial coefficients~$a_k(N)$.

There is no general proof that the $k$-loop contributions to a 1-scale
observable have to obey such a recurrence. On the other hand, it is known that
all single scale processes having been calculated so far do,
cf. \cite{X2,COHS,MVV1,MVV2,MVV3}.
This is due to the fact that the corresponding observables are found as
linear combinations of nested harmonic sums. The single harmonic sums obey
%----------------------------------------------------------------------------
\begin{equation*}
F(N+1) - F(N) = \frac{{\rm sign}(a)^{N+1}}{(N+1)^{|a|}}.
\end{equation*}
%----------------------------------------------------------------------------
Exploiting holonomic closure properties~\cite{salvy:94} one obtains higher
order difference equations for polynomial expressions in terms of nested harmonic sums.

If a suitably large number of moments $\Mvec[f(x)](N)$ is known,
then a recurrence of the form~\eqref{eqDI} can
be found automatically, see Section~\ref{sec:2}.
Once a recurrence of some order~$l$ is found, this recurrence together with the first $l$~moments
specifies uniquely all the moments $\Mvec[f(x)](N)$ for nonnegative integers~$N$.
Finally, we activate the summation package~\SigmaP~\cite{Schneider:07a} and solve the recurrence~\eqref{eqDI} in terms of generalized harmonic sums.
In particular, using the underlying summation theory of $\Pi\Sigma$-difference
fields~\cite{Karr:81,Schneider:08d,Schneider:08e} or exploiting the algebraic
relations~\cite{ALGEBRA},
a closed form for $\Mvec[f(x)](N)$ in terms of an algebraically independent basis of harmonic sums can be computed.

We emphasize that
Eq.~(\ref{eqDI}) covers a much wider class in which more general
recurrent quantities can represent the corresponding observables. In particular, our general recurrence solver for d'Alembertian
solutions~\cite{Noerlund,Abramov:94,Schneider:01a} finds any solution that can be expressed in terms of indefinite nested sums and products. In even higher order or massive calculations further
functions may contribute, which could be only found in this way.

The Mellin-moments of the unpolarized 3--loop splitting functions and Wilson
coefficients for deep--inelastic scattering are more easily calculated
\cite{LRV,MOM,MOM1,HMOM} than the complete expressions,
cf.~\cite{MVV1,MVV2,MVV3}. In the present paper
we investigate whether the exact formulae up to  the unpolarized 3--loop
anomalous dimensions and Wilson coefficients \cite{MVV1,MVV2,MVV3} can be found
establishing and solving difference equations \eqref{eqDI} for the Mellin moments of
these quantities, without further assumptions.\footnote{Approximate reconstruction
methods based on special ansatzes were discussed in the literature e.g. in
\cite{LRV,NV} to obtain first numerical estimates from a low number of moments, see
also~\cite{HARKIL}. We also remind that the description of QCD-evolution relating
fixed integer moment information to orthogonal polynomials is an old topic
\cite{ORTHPOL}; see also \cite{GROSS}.}

We consider the various color contributions to these quantities separately and try to
find the complete result from a minimal number of moments. As input we apply the moments
calculated from the exact solution~\cite{MVV1,MVV2,MVV3}.

The paper is organized as follows. In Section~\ref{sec:2} we describe how the difference
equations of the form~\eqref{eqDI} are found by just using a finite number of starting
points of~$F(N)$. In Section~\ref{sec:3} the algorithms are outlined that can solve
these recurrences in the setting of difference fields. They lead
directly to the corresponding mathematical structures. These are nested
harmonic sums in the present case. In course of the solution we compactify
the results applying the algebraic relations to the harmonic
sums~\cite{ALGEBRA}.~\footnote{Further compactifications can be obtained using the
structural relations, cf.~\cite{STRUCT5,STRUCT6}.} The results are discussed
in Section~\ref{sec:4}.
Our method applies in the same way to all other single scale processes of similar
complexity, cf. \cite{COHS,HEAV}. Section~\ref{sec:5} contains the conclusions. In
the appendix we present a compactified form of the non-singlet
3--loop anomalous dimensions, which is automatically provided in the formalism
by~\SigmaP. The corresponding expressions for the other anomalous
dimensions and Wilson coefficients to 3-loop order are presented in
\mathP\ and \FORMP\ codes attached.
%%%%%%%%%%%%%%%%%%%%%%%%%%%%%%%%%%%%%%%%%%%%%%%%%%%%%%%%%%%%%%%%%%%%%%%
\section{Finding a Recurrence Equation}\label{sec:2}
%%%%%%%%%%%%%%%%%%%%%%%%%%%%%%%%%%%%%%%%%%%%%%%%%%%%%%%%%%%%%%%%%%%%%%%

Suppose we are given a finite array of rational numbers,
%------------------------------------------------------------------------------------
\begin{alignat*}1
  &q_1,\ q_2,\ \dots,\ q_K~,
\end{alignat*}
%------------------------------------------------------------------------------------
 which are the first terms of a certain infinite
 sequence~$F(N)$,~i.e., $F(1)=q_1$, $F(2)=q_2$, etc.
 Let us assume that $F(N)$
 satisfies a recurrence of type
%------------------------------------------------------------------------------------
\begin{eqnarray}
\label{eq:rec}
  \sum_{k=0}^l\Bigl(\sum_{i=0}^d c_{i,k} N^i\Bigr)F(N+k)=0~,
\end{eqnarray}
%------------------------------------------------------------------------------------
 which
 we would like to deduce from the given numbers~$q_i$ ($i=1,\dots,K$). In a
 strict sense, this is not possible without knowing how the sequence
 continues for $N>K$. One thing we can do is to determine the recurrence
 equations satisfied by the data we are given. Any recurrence for $F(N)$ must
 certainly be among those.

 To find the recurrence equations of $F(N)$ valid for the first terms, the
 simplest way to proceed is by making an ansatz with undetermined coefficients.
 Let us
 fix an order~$l\in\mathbf{N}$ and a degree~$d\in\mathbf{N}$ and consider
 the generic recurrence (\ref{eq:rec}),
 where the $c_{i,k}$ are indeterminates. For each specific choice $N=1,2,\dots,K-l$,
 we can evaluate the ansatz, because we know all the values of $F(N+k)$ in
 this range, and we obtain a system of $K-l$ homogeneous linear equations
 for $(l+1)(d+1)$ unknowns~$c_{i,j}$.

 If $K-l>(l+1)(d+1)$, this system is under-determined and is thus guaranteed
 to have nontrivial solutions. All these solutions will be valid recurrences
 for $F(N)$ for $N=1,\dots,K-l$, but they will most typically fail to hold
 beyond. If, on the other hand, $K-l\leq(l+1)(d+1)$, then the system is
 overdetermined and nontrivial solutions are not to be expected. But
 at least recurrence equations valid for all~$N$, if there are any, must
 appear among the solutions. We therefore expect in this case that the
 solution set will precisely consist of the recurrences of~$F(N)$
 of order~$l$ and degree~$d$ valid for all~$N$.

 % example

 As an example, let us consider the contribution to the gluon splitting function
$\propto C_A$ at leading order,
$P_{gg}^{(0)}(N)$. The first 20 terms, starting with $N=3$, of the sequence $F(N)$ are
%------------------------------------------------------------------------------------
 \begin{alignat*}1
   &\tfrac{14}{5},\ \tfrac{21}{5},\ \tfrac{181}{35},\ \tfrac{83}{14},\
   \tfrac{4129}{630},\ \tfrac{319}{45},\ \tfrac{26186}{3465},\
   \tfrac{18421}{2310},\ \tfrac{752327}{90090},\ \tfrac{71203}{8190},\
   \tfrac{811637}{90090},\ \tfrac{128911}{13860},\
   \tfrac{29321129}{3063060},\\ &
   \tfrac{2508266}{255255},\ \tfrac{292886261}{29099070},\
   \tfrac{7045513}{684684},\ \tfrac{611259269}{58198140},\
   \tfrac{1561447}{145860},\ \tfrac{4862237357}{446185740},\
   \tfrac{988808455}{89237148}~.
 \end{alignat*}
%------------------------------------------------------------------------------------
 Making an ansatz for a recurrence of order~3 with polynomial coefficients
 of degree~3 leads to an overdetermined homogeneous linear system
 with 16 unknowns and 17 equations. Despite of being overdetermined and dense,
 this system has two linearly independent solutions. Using bounds for the
 absolute value of determinants depending on the size of a matrix and the
 bit size of its coefficients, one can very roughly estimate the probability
 for this to happen ``by coincidence''
 to about~$10^{-65}$. And in fact, it did not happen
 by coincidence. The solutions to the system correspond to the two
 recurrence equations
%------------------------------------------------------------------------------------
 \begin{alignat}1
  &(7 N^3+113 N^2+494 N+592) F(N)-(12 N^3+233 N^2+1289 N+2156) F(N+1)\notag\\
  &{}+(3 N^3+118 N^2+1021 N+2476) F(N+2)+(2 N^3+2 N^2-226 N-912) F(N+3)=0\label{eq1}
 \end{alignat}
 and
 \begin{alignat}1
  &(4 N^3+64 N^2+278 N+332) F(N)-(7 N^3+134 N^2+735N+1222) F(N+1)\notag\\
  &{}+(2 N^3+71 N^2+595 N+1418) F(N+2)+(N^3-N^2-138 N-528) F(N+3)=0,\label{eq2}
 \end{alignat}
%------------------------------------------------------------------------------------
 which both are valid for all~$N\geq1$. If we had found that the linear system
 did not have a nontrivial solution, then we could have concluded that the
 sequence $F(N)$ would \emph{definitely} (i.e.\ without any uncertainty)
 not satisfy a recurrence of order~3 and degree~3.
 It might then still have satisfied recurrences with larger order or degree,
 but more terms of the sequence had to be known for detecting those.

 % references, history

 The method of determining (potential) recurrence equations for sequences
 as just described is not new.
 It is known to the experimental mathematics community as
 \emph{automated guessing} and is frequently applied in the study of
 combinatorial sequences. Standard software packages for generating functions
 such as \textsf{gfun}~\cite{salvy:94} for \textsf{Maple} or
 \textsf{GeneratingFunctions.m}~\cite{mallinger96} for
 \mathP\ provide functions which take as input a finite array of numbers,
 thought of as the first terms of some infinite sequence, and produce as output
 recurrence equations that are, with high probability, satisfied by the
 infinite sequence.

 % modular approach

 These packages apply the method described above more or less literally,
 and this is perfectly sufficient for small examples. But if thousands of
 terms of a sequence are needed, there is no way to get the linear
 systems solved using rational number arithmetic. Even worse, already for medium
 sized problems from our collection, the size of the linear system exceeds by
 far typical memory capacities of~16--64Gb. For the big
 problem~$C^{(3)}_{2,q,C_F^3}(N)$,
 it would require approximately 11Tb of memory to represent the corresponding linear
 system explicitly. It is thus evident that computations with rational numbers
 are not feasible. Instead, we use arithmetic in finite fields together with
 Chinese remaindering and rational
 reconstruction~\cite{geddes92,vzgathen99,kauers08j}. Modulo a word
 size prime, the size of the biggest systems reduces to a few Gb, a size
 which easily fits on our architecture. And modulo a word size prime, such
 a system can be solved within no more than a few hours of computation time by
 \mathP.

 The modular results for several distinct primes $p_1,p_2,\dots$ can be combined
 by Chinese remaindering to a modular result whose coefficients are correct
 modulo the product $p_1p_2\dots$. If the bit size of this product exceeds
 twice the maximum bit size appearing in the rational solution, then the
 exact rational number coefficients can be recovered from the modular images
 by rational reconstruction~\cite{geddes92,vzgathen99,kauers08j}.
 The number of primes needed (and thus
 the overall runtime) is therefore proportional to the bit size of the coefficients
 in the final output.

 % operator gcds: explain by continuing example
 The final output is a recurrence equation for~$F(N)$.
 But the recurrence equation satisfied by a sequence $F(N)$ is not unique: if
 a sequence satisfies a recurrence equation at all, then it satisfies a
 variety of linearly independent recurrence equations.
 The bit size of the rational number coefficients in these recurrence equations
 may vary dramatically.
 In order to minimize the number of primes needed for the computation of the
 rational numbers in the recurrence, it seems preferable to compute on a recurrence
 whose coefficients are as small as possible in terms of bit size.
 According to our experience, this recurrence happens to be the (unique) recurrence
 whose order~$l$ is minimal among all the recurrence equations satisfied by~$F(N)$.
 We have no explanation for this, but it seems to be a general phenomenon, as it
 can also be observed in certain combinatorial applications~\cite{bostan08}.

 Also the number of unknowns for the linear system may vary
 dramatically among the possible recurrence equations for~$F(N)$, and
 it seems preferable to compute on a recurrence where the number of unknowns is
 as small as possible. Small linear systems are not only preferable because of
 efficiency, but also because the number of unknowns in the linear system
 determines the number of initial terms $q_i$ that have to be known a priori
 in order to detect the recurrence. According to our experience, the size of the
 linear system is minimized when the order~$l$ and the degree~$d$ are approximately
 balanced.

 Unfortunately, it seems that the recurrence with \emph{minimal} (in terms of
 bit size) rational number coefficients has the \emph{maximal} number of unknowns
 in the corresponding linear system, and vice versa.
 But there is a way to combine the advantages of both at a reasonable computational
 cost. Consider the
 two recurrence equations \eqref{eq1} and \eqref{eq2} from the example
 of the gluon--gluon splitting function at leading order, $P_{gg,0}(N)$,
 quoted above.
 A recurrence
 of smaller order can be obtained from these by multiplying \eqref{eq1}
 by $(N^2 - 9 N - 66)$ and \eqref{eq2} by $(2 N^2 - 14 N - 114)$, and then
subtracting the results.
 The choice of the multipliers is such that the coefficient of $F(N+3)$ in the
 difference cancels: we obtain
\begin{alignat*}1
&(N^5+22 N^4+189 N^3+788 N^2+1592 N+1224) F(N)\\
&{} -(2 N^5+45 N^4+396N^3+1701 N^2+3580 N+2988) F(N+1)\\
&{} +(N^5+23 N^4+207 N^3+913 N^2+1988 N+1764) F(N+2) = 0.
 \end{alignat*}
 The calculation just performed can be recognized as the first step in a difference
 operator version of the Euclidean algorithm~\cite{bronstein96}.
 Applied to two recurrence
 equations satisfied by
 a sequence~$F(N)$, this algorithm yields their ``greatest common (right) divisor'',
 which is, with high probability, the minimal
 order recurrence satisfied by~$F(N)$. In our example, the algorithm terminates
 in the next step, and indeed the sequence $F(N)$ of $P_{gg,0}(N)$ does not
 satisfy a recurrence of order less than two. Note that the linear system
 for finding the second order recurrence directly would have involved $(5+1)(2+1)=18$
 unknowns instead of the 16 unknowns we needed for finding the third order recurrences.
 For the big problem~$C^{(3)}_{2,q,C_F^3}(N)$, a direct computation would require 33804
 unknowns instead of the 5022 we actually used.
 We combine the advantage of a small linear system with the advantage of small
 coefficients in the output as follows. We first compute for several word size primes
 the solutions of a small linear system, but then instead of applying rational
 reconstruction to those, we compute, for each prime independently, their
 greatest common right divisor modulo this prime. We then apply rational
 reconstruction to recover the rational number coefficients of those.

 % summarize 'algorithm'.
 In summary, we used the following procedure for finding the recurrence equations.
 \begin{enumerate}
 \item\label{step:1} Choose a word size prime~$p$.
 \item\label{step:2} Choose some bounds $l$ and $d$ and make an ansatz for a recurrence
   of order~$l$ and degree~$d$. The linear system is constructed and solved
   modulo~$p$ only.
 \item\label{step:3} If there are no solutions, repeat step~\ref{step:2} with increased
   bounds $l$ and~$d$.
 \item\label{step:4} If there are solutions modulo~$p$,
   compute their greatest common right divisor modulo~$p$
   by the Euclidean algorithm for difference operators.
 \item\label{step:5} Repeat steps \ref{step:1}--\ref{step:4} until Chinese
   remaindering and rational reconstruction applied to the greatest common right
   divisors for the various primes yields a recurrence that matches the given data
   $q_1,q_2,\dots,q_K$.
 \item\label{step:6} Return the reconstructed recurrence as the final result.
 \end{enumerate}

 % remarks on the large scale computations.
 For the big problem~$C^{(3)}_{2,q,C_F^3}(N)$, most of the computation time (about 53\%)
was
 spent in step~\ref{step:4}. Solving the modular linear systems consumed about 28\%
 of the time, and Chinese remaindering and rational reconstruction took about 18\%
 of the time. The memory bottleneck is in step~\ref{step:2} where the linear system
 is constructed. The memory requirements for the other steps, if implemented well,
 are negligible.

 % potential further improvements:
 % hermite pade, platforms other than mathematica, parallel computations
 For problems that are even bigger than those we considered, further improvements
 to the procedure are conceivable. First, there are asymptotically fast special
 purpose algorithms for step~\ref{step:2}
available~\cite{beckermann92,beckermann00}.
 These algorithms
 outperform the naive linear system approach we are taking for problem sizes where
 fast polynomial multiplication algorithms outperform classical algorithms. It is
 likely that their use would have already been beneficial for some of our problems.
 Second, a gain in efficiency might result from running the procedure on a different
 platform. We have done all our computations within \mathP~6, but we
expect that
 in particular step~\ref{step:4} might considerably benefit from a reimplementation
 in a computer algebra system providing high-performance polynomial arithmetic.
 \mathP's modular arithmetic, on the other hand, appears to be quite
 competitive.
 Third, it might be worthwhile to run parts of the procedure in parallel.
 In particular, computations for distinct primes are completely independent from
 each other and can be done on different processors without any communication
 overhead. Observe that these steps dominate the runtime.
%%%%%%%%%%%%%%%%%%%%%%%%%%%%%%%%%%%%%%%%%%%%%%%%%%%%%%%%%%%%%%%%%%%%%%%%%%%%%%%%%%%%%%%%
\section{Solving the Recurrence Equations}\label{sec:3}
%%%%%%%%%%%%%%%%%%%%%%%%%%%%%%%%%%%%%%%%%%%%%%%%%%%%%%%%%%%%%%%%%%%%%%%%%%%%%%%%%%%%%%%%

After having obtained difference equations of high order and degree we will now
discuss general, efficient algorithms by which these equations can be solved.
\textit{Given} a recurrence relation
%-------------------------------------------------------------------------------------
\begin{equation}
\label{Equ:GeneralRec}
a_0(N)F(N)+a_1(N)F(N+1)+\dots+a_l(N)F(N+l)=q(N)
\end{equation}
%-------------------------------------------------------------------------------------
of order $l$, \textit{find} all its solutions that can be expressed in terms of
indefinite nested sums and products. Such solutions are also called d'Alembertian
solutions~\cite{Noerlund,Abramov:94,Schneider:01a}, they form a subclass of Liouvillian
solutions~\cite{Singer:99}. Note that such solutions cover as special cases, e.g.,
harmonic sums~\cite{HSUM,summer} or generalized nested harmonic
sums~\cite{Moch:02}.

The solution to this problem consists of two parts.
%-------------------------------------------------------------------------------------
\begin{enumerate}
\item
First, compute all d'Alembertian solutions by factoring the recurrence
as much as possible into linear right factors. Then each linear factor contributes
to one extra solution. To be more precise, the $i$th factor yields a nested sum
expression of depth $i-1$.
\item
Second, simplify these nested sum solutions to closed form expressions, e.g., in terms
of harmonic sums, that can be processed further in practical problem solving.
\end{enumerate}
%-------------------------------------------------------------------------------------

In general, the package \SigmaP~\cite{Schneider:07a} can solve these problems in the
setting of $\Pi\Sigma$-difference fields~\cite{Karr:81,Karr:85}. This means that the
coefficients $a_0(N),\dots,a_l(N)$ and the inhomogeneous part $q(N)$
of~\eqref{Equ:GeneralRec} can be given as polynomial expressions in terms of
indefinite nested sums and products.

For simplicity, we restrict ourself to the situation that the given coefficients
$a_0(N),\dots,a_l(N)$ are polynomials in $N$ and that the inhomogeneous part $q(N)$
is zero. In other words, we assume that we are given a recurrence of the
form~\eqref{eqDI} or~\eqref{eq:rec} that is produced, e.g., by the method described in the
previous section.

%%%%%%%%%%%%%%%%%%%%%%%%%%%%%%%%%%%%%%%%%%%%%%%%%%%%%%%%%%%%%%%%%%%%%%%%%%%%%%%%%%%%%%%%
\subsection{Finding all d'Alembertian solutions}
%%%%%%%%%%%%%%%%%%%%%%%%%%%%%%%%%%%%%%%%%%%%%%%%%%%%%%%%%%%%%%%%%%%%%%%%%%%%%%%%%%%%%%%%

Subsequently, we present algorithms that find all d'Alembertian solutions
of~\eqref{eqDI}. Equivalently, we can say that we look for all
d'Alembertian sequences which are annihilated by the linear operator
%-------------------------------------------------------------------------------------
\begin{equation}\label{Equ:LOp}
L:=a_0(N)+a_1(N)\SE+\dots+a_l(N)\SE^l,
\end{equation}
%-------------------------------------------------------------------------------------
which is understood to act on a sequence $F(N)$ via
\[
 (L\cdot F)(N):=a_0(N)F(N)+a_1(N)F(N+1)+\dots+a_l(N)F(N+l).
\]
We start as follows.

\medskip
\noindent\textit{Step 1: Finding a product solution.} First, we look for a solution of~\eqref{Equ:LOp} which is of the form
\begin{equation}\label{ProdSol}
T_0(N)=\prod_{i=\lambda}^Nr(i)
\end{equation}
for some rational function $r(i)$ in $i$. In \SigmaP\ this task can be carried out by executing a generalized version of algorithm~\cite{Petkov:92} that works in general $\Pi\Sigma$-difference fields; for an alternative algorithm to find such hypergeometric terms we refer to~\cite{vanHoeij:99}.

If there does not exist such a product solution~\eqref{ProdSol}, then there is
no d'Alembertian solution at all; see, e.g.~\cite[Theorem~4.5.5]{Schneider:01a}. In
this case, we just stop. Otherwise, we look for additional solutions as follows.

\medskip
\noindent\textit{Step 2: Splitting off a linear right factor.} By dividing the
operator~\eqref{Equ:LOp} from the right with the operator
%-------------------------------------------------------------------------------------
\begin{equation}
\SE-\frac{T_0(N+1)}{T_0(N)}=\SE-r(N+1)
\end{equation}
%-------------------------------------------------------------------------------------
we arrive at an operator
%-------------------------------------------------------------------------------------
\begin{equation}\label{Equ:OperatorL}
L':=b_0(N)+b_1(N)\SE+\dots+b_{l-1}(N)\SE^{l-1}
\end{equation}
%-------------------------------------------------------------------------------------
of order $l-1$ such that
%-------------------------------------------------------------------------------------
\begin{align*}
L&=L'(\SE-r(N+1))\\
&=-r(N+1)b_0(N)+\big(b_0(N)-r(N+2)b_1(N)\big)\SE
+\dots+\big(b_{l-1}(N)-r(N+l)b_l(N)\big)\SE^{l},
\end{align*}
%-------------------------------------------------------------------------------------
i.e., $\SE-r(N+1)$ is a linear right factor of $L$.

\medskip
\noindent\textit{Step 3: Recursion.} Now we continue by recursion and look for all
d'Alembertian solutions for the operator $L'$ with order $l-1$. Note that after at most $l-1$ steps we end up at a recurrence of order $1$ whose d'Alembertian solution
can be read off immediately.

\medskip
\noindent\textit{Step 4: Combining the solutions.} If we do not find any
d'Alembertian solution for $L'$, we just return the solution~\eqref{ProdSol} for $L$.\\
Otherwise, let
%-------------------------------------------------------------------------------------
\begin{equation}\label{Equ:dAlembSolSub}
t_1(N),\dots,t_k(N)
\end{equation}
%-------------------------------------------------------------------------------------
with $1\leq k<l$ be the solutions of $L'$ that we obtained after the recursion step.
To this end, for $1\leq j\leq k$ define
%-------------------------------------------------------------------------------------
\begin{equation}
\label{Equ:Sj}
T_j(N):=T_0(N)\sum_{i=\lambda}^N\frac{t_j(i-1)}{T_0(i)}
\end{equation}
%-------------------------------------------------------------------------------------
for some properly chosen $\lambda\geq0$ (i.e., $T_0(i)$ is nonzero for all $i$ with
$i\geq\lambda$). Then the final output of our algorithm is
%-------------------------------------------------------------------------------------
\begin{equation}
\label{Equ:dAlembSol}
T_0(N),T_1(N),\dots,T_k(N).
\end{equation}
%-------------------------------------------------------------------------------------
The following remarks are in place. By construction all the elements
from~\eqref{Equ:dAlembSol} are solutions of~\eqref{Equ:LOp}: for each $1\leq j<k$,
%-------------------------------------------------------------------------------------
\begin{align*}
(\SE-&r(N+1))\cdot T_j(N)=
T_0(N+1)\sum_{i=\lambda}^{N+1}\frac{t_j(i-1)}{T_0(i)}-r(N+1)T_0(N)\sum_{i=\lambda}^N\frac{t_j(i-1)}{T_0(i)}\\
&=r(N+1)T_0(N)\left(\sum_{i=\lambda}^{N}\frac{t_j(i-1)}{T_0(i)}+\frac{t_j(N)}{T_0(N+1)}\right)-r(N+1)T_0(N)\sum_{i=\lambda}^N\frac{t_j(i-1)}{T_0(i)}=t_j(N)
\end{align*}
%-------------------------------------------------------------------------------------
and hence
$$L\cdot T_j(N)=L'\cdot t_j(N)=0.$$
But even more holds. The derived solutions~\eqref{Equ:dAlembSol} are linearly
independent. In particular, any solution of $L$ in terms of indefinite nested
sums and products can be expressed as a linear combination of~\eqref{Equ:dAlembSol};
see~\cite[Theorem~5.1]{Singer:99} or~\cite[Proposition~4.5.2]{Schneider:01a}.

\smallskip

Summarizing, with the algorithm sketched above we can produce all d'Alembertian
solutions of $L$, i.e., all solutions that are expressible in terms of indefinite
nested sums and products.

\medskip

We emphasize that the expensive part of the sketched method is the computation of
the product solutions~\eqref{ProdSol}.
The following improvements were crucial in order to solve the recurrences under
consideration.

\smallskip

\noindent\textit{Improvement 1.} If one finds several product solutions,
say $P_1(N),\dots,P_u(N)$, one can produce immediately a recurrence $L'$
like in~\eqref{Equ:OperatorL}, but with order $l-u$ instead of order $l-1$.
Moreover, given all d'Alembertian solutions of this operator $L'$, one gets all
the solutions of the recurrence~\eqref{Equ:LOp} without any further computations;
see~\cite[Theorem~4.5.6]{Schneider:01a}.

\smallskip

\noindent\textit{Improvement 2.}
For the problems under consideration, it turns out that it suffices to search for
product solutions~\eqref{ProdSol} that can be written in the form
%-------------------------------------------------------------------------------------
\begin{equation}
\label{Equ:PForm}
T_0(N)=\frac{p(N)}{q(N)}\quad\text{or}\quad T_0(N)=\frac{p(N)}{q(N)}(-1)^N
\end{equation}
%-------------------------------------------------------------------------------------
for polynomials $p(N)$ and $q(N)$. Therefore, we used optimized
solvers~\cite{Schneider:05a} of \SigmaP\ which generalize the algorithm presented
in~\cite{Abramov:71}. In addition, arithmetic in finite fields is exploited in order
to determine the solutions~\eqref{Equ:PForm} effectively.

\smallskip

\noindent\textit{Improvement 3.} In our applications, rather big factors
from $t_j(i-1)$ and $T_0(i)$ cancel in the summand
$\frac{t_j(i-1)}{T_0(i)}$ of~\eqref{Equ:Sj}; in particular, the usually irreducible
factor $p(i)$ from~\eqref{Equ:PForm} ($N$ substituted with $i$) cancels. Hence it
pays off to compute directly the summand expression $\frac{t_j(i-1)}{T_0(i)}$: Namely,
instead of the operator~\eqref{Equ:OperatorL} we continue with the operator
%-------------------------------------------------------------------------------------
\begin{equation}\label{Equ:SubMod}
b_0(N)+r(N)b_1(N)\SE+\dots+\left(\prod_{i=1}^{l-1}r(N+i)\right)b_{l-1}(N)\SE^{l-1},
\end{equation}
%-------------------------------------------------------------------------------------
and look for all its d'Alembertian solutions
$t'_1(N),\dots,t'_k(N)$.
%of the recurrence
%\begin{equation}\label{Equ:SubMod}
%b_0(N-1)G(N)+r(N-1)b_1(N-1)G(N)+\dots+\left(\prod_{i=1}^{l-1}r(N+i-1)\right)b_{l-1}(N-1)G(N+l-1)=0.
%\end{equation}
Then by construction, the solutions of~\eqref{Equ:GeneralRec} can be given directly
in the form
%-------------------------------------------------------------------------------------
\begin{equation}\label{Equ:SjImproved}
T_j(N)=T_0(N)\sum_{i=\lambda}^Nt'_j(i).
\end{equation}
%-------------------------------------------------------------------------------------

\begin{example}
As illustrative example we solve the difference equation for the $C_F 
N_F^2$-term  of
the unpolarized 3-loop splitting function $P_{gq,2}(N)$,
%-------------------------------------------------------------------------------------
\begin{equation}
F(N)=P_{gq,2,N_F^2C_F}(N)~.
\nonumber
\end{equation}
%-------------------------------------------------------------------------------------
Using the methods from the previous section, we generate the recurrence relation
%-------------------------------------------------------------------------------------
\begin{eqnarray}
\label{Exp:Rec3}
a_0(N) F(N)
+a_1(N) F(N+1)
+a_2(N) F(N+2)
+a_3(N) F(N+3) = 0~,
\end{eqnarray}
with
%-------------------------------------------------------------------------------------
\begin{alignat*}1
a_0(N) &= (1-N) N (N+1) (N^6+15 N^5+109 N^4+485 N^3+1358 N^2+2216 N+1616),\\
a_1(N) &= N (N+1)(3 N^7+48 N^6+366 N^5+1740 N^4+5527 N^3+11576 N^2+14652N+8592),\\
a_2(N) &= -(N+1)(3 N^8+54 N^7+457 N^6+2441 N^5+9064 N^4+23613 N^3\\
       &\qquad{}+41180 N^2+43172 N+20768),\\
a_3(N) &= (N+4)^3(N^6+9 N^5+49 N^4+179 N^3+422 N^2+588 N+368)~.
\end{alignat*}
%-------------------------------------------------------------------------------------
Given this recurrence, we produce its d'Alembertian solutions as follows.
First, \SigmaP\ computes a rational solution, namely
$$T_0(N)=\frac{N^2+N+2}{(N-1)N(N+1)}.$$
Now we can divide~\eqref{Exp:Rec3} from the right by the operator
$\SE-T_0(N+1)/T_0(N)=S-\frac{(N-1) \left(N^2+3 N+4\right)}{(N+2) \left(N^2+N+2\right)}.$
Then the resulting recurrence of the operator~\eqref{Equ:SubMod} is
%-------------------------------------------------------------------------------------
\begin{eqnarray}
\label{Exp:Rec2}
 b_0(N) G(N)
+b_1(N) G(N+1)
+b_2(N) G(N+2) = 0~,
\end{eqnarray}
%-------------------------------------------------------------------------------------
with
%-------------------------------------------------------------------------------------
\begin{alignat*}1
b_0(N) &= (N+1)(N+2)(N^2-N+2)(N^6+9 N^5+49 N^4+179 N^3+422 N^2+588 N+368),\\
b_1(N) &= -(N+2)(2 N^9+21 N^8+124 N^7+530 N^6+1690 N^5+3989N^4+6712 N^3+7524 N^2\\
       &\qquad{}+5232 N+2080),\\
b_2(N) &= (N^2+5 N+8) (N^6+3 N^5+19 N^4+53 N^3+104 N^2+124N+64) (N+3)^2~.
\end{alignat*}
%-------------------------------------------------------------------------------------
Next, we proceed recursively and can compute the rational solution
$$P'(N)=\frac{N^4+4 N^3+13 N^2+22 N+16}{(N-1) N (N+1) (N+2) \left(N^2+3 N+4\right)}$$
of~\eqref{Exp:Rec2}. Thus we divide~\eqref{Exp:Rec2} by the factor $\SE-\frac{P'(N+1)}{P'(N)}$
%$$\SE-\frac{P'(N+1)}{P'(N)}=S-\frac{(N+2) \left(N^2+N+2\right) %\left(N^4+8 N^3+31 N^2+64
%   N+56\right)}{(N+3) \left(N^2+5 N+8\right) \left(N^4+4 N^3+13 %N^2+22
%   N+16\right)}$$
which leads to the first order recurrence
%-------------------------------------------------------------------------------------
\begin{eqnarray}
c_0(N) H(N) + c_1(N) H(N+1) = 0
\end{eqnarray}
%-------------------------------------------------------------------------------------
with
%-------------------------------------------------------------------------------------
\begin{alignat*}1
c_0(N) &=-(N+1)(N^2+N+2)(N^4-4 N^3+13 N^2-14 N+8)\\
  &\qquad{}\times\left(N^6+3 N^5+19 N^4+53 N^3+104 N^2+124 N+64\right),\\
c_1(N) &= (N+2)(N^2-N+2)(N^4+4 N^3+13 N^2+22 N+16)\\
  &\qquad{}\times(N^6-3 N^5+19 N^4-13 N^3+44 N^2+8 N+8).
\end{alignat*}
Here we can read off directly the solution
%-------------------------------------------------------------------------------------
\begin{eqnarray*}
P''(N)=\frac{\left(N^2-N+2\right) \left(N^6-3 N^5+19 N^4-13 N^3+44 N^2+8
   N+8\right)}{(N+1) \left(N^4+7 N^2+4 N+4\right) \left(N^4-4 N^3+13 N^2-14
   N+8\right)}.
\end{eqnarray*}
%-------------------------------------------------------------------------------------
Going back, we obtain besides $t_1(N)=P'(N)$ the solution
\begin{align*}
t_2(N)&=P'(N)\sum_{j=1}^NP''(j)=\tfrac{N^4+7
   N^2+4 N+4}{(N+1) \left(N^2-N+2\right)
   \left(N^2+N+2\right)} \sum_{j=1}^N\tfrac{\left(j^2-j+2\right)
   \left(j^6-3 j^5+19 j^4-13 j^3+44 j^2+8 j+8\right)}{(j+1)
   \left(j^4+7 j^2+4 j+4\right) \left(j^4-4 j^3+13 j^2-14
   j+8\right)}
\end{align*}
of~\eqref{Exp:Rec2}. Hence by~\eqref{Equ:SjImproved} we obtain, besides $T_0(N)$, the solutions
\begin{equation}\label{Equ:DoubleSumSol}
\begin{split}
 T_1(N)= & \frac{\displaystyle\left(N^2+N+2\right) \sum_{i=1}^N\frac{i^4+7 i^2+4
   i+4}{(i+1) \left(i^2-i+2\right)
   \left(i^2+i+2\right)}}{(N-1) N (N+1)},\\
 T_2(n)= & \frac{\displaystyle\left(N^2+N+2\right) \sum_{i=1}^N\frac{\displaystyle\left(i^4+7
   i^2+4 i+4\right) \sum_{j=1}^i\tfrac{\left(j^2-j+2\right)
   \left(j^6-3 j^5+19 j^4-13 j^3+44 j^2+8 j+8\right)}{(j+1)
   \left(j^4+7 j^2+4 j+4\right) \left(j^4-4 j^3+13 j^2-14
   j+8\right)}}{(i+1) \left(i^2-i+2\right)
   \left(i^2+i+2\right)}}{(N-1) N (N+1)}
\end{split}
\end{equation}
for~\eqref{Exp:Rec3}. Since all three solutions $T_0(N)$, $T_1(N)$ and $T_2(N)$ are linearly independent over, say, the complex numbers, any solution $F:{\mathbf{N}}\to{\mathbf{C}}$ of~\eqref{Exp:Rec3} can be described as a linear combination
$$F(N)=c_1T_0(N)+c_2T_1(N)+c_3T_2(N)$$
for $c_1,c_2,c_3\in\mathbf{C}$. The initial values
$F(3)=\tfrac{1267}{648},F(4)=\tfrac{54731}{40500},F(5)=\tfrac{20729}{20250}$
imply
\begin{equation}\label{Equ:Solution2}
P_{gq,2,N_F^2C_F}(N)=-\tfrac{32}{9}T_0(N)+\tfrac{64}{9}T_1(N)-\tfrac{8}{3}T_2(N).
\end{equation}
\end{example}

For our concrete problems all the recurrences could be factored completely. Equivalently, for a recurrence of order $d$ we found $d$ linearly independent solutions $T_1(N),\dots,T_d(N)$ where the solution $T_k$ with $1\leq k\leq d$ can be given in the form
\begin{equation}\label{Equ:dAlembertSol}
T_k(N)=s_0^N\frac{P_0(N)}{Q_0(N)}\sum_{i_1=1}^Ns_1^{i_1}\frac{P_1(i_1)}{Q_1(i_1)}\sum_{i_2=1}^{i_1}s_2^{i_2}\frac{P_2(i_2)}{Q_2(i_2)}\dots\sum_{i_k=1}^{i_{k-1}}s_k^{i_k}\frac{P_k(i_k)}{Q_k(i_k)}
\end{equation}
where for $1\leq i\leq k$ the $P_i$ and $Q_i$ are polynomials and $s_i\in\{-1,1\}$.

\begin{example}\label{Exp:7Order}
For the $C_A C_F N_F$-term of the 3-loop non-singlet splitting function 
$P_{NS,2}^{-}$
we found a recurrence of order~7 which fills
around five pages. The 7~linearly independent solutions can be computed within 10~seconds; the largest solution fills around three pages and has the form
\begin{equation}\label{Equ:Solution1}
\sum_{i=6}^N\frac{P_1(i)}{Q_1(i)}\sum_{j=1}^i\frac{P_2(j)}{Q_2(j)}\sum_{k=5}^j\frac{P_3(k)}{Q_3(k)}\sum_{l=1}^k\frac{P_4(l)}{Q_4(l)}\sum_{r=1}^l\frac{P_5(r)}{Q_5(r)}\sum_{s=2}^r\frac{P_6(s)}{Q_6(s)}
\end{equation}
where the irreducible polynomials $P_1,P_2,\dots,P_6$ have the respective degrees
4,~8, 16, 28, 63,~69,
and the denominators are of the form
{\allowdisplaybreaks\small
\begin{align*}
Q_1(s)=&(s-1)^3 s (s+1)^3,\\
Q_2(r)=&\left(r^4-10 r^3+29 r^2-34 r+12\right) \left(r^4-6 r^3+5 r^2-2r-2\right),\\
Q_3(l)=&\big(l^8-24 l^7+215 l^6-1017 l^5+2866 l^4-4975 l^3+5146 l^2-2812
   l+576\big)\\
   &\times\big(l^8-16 l^7+75 l^6-175 l^5+236 l^4-165 l^3-4 l^2+64
   l-24\big),\\
Q_4(k)=&(k-4) (k-3)^2 (k-2)^3 (k-1)^3 k^2 \big(k^{10}-38 k^9+566 k^8-4628
   k^7+23621 k^6-79466 k^5\\
   &+178404 k^4-261580 k^3+235712 k^2-114624
   k+21600\big) \big(k^{10}-28 k^9+269 k^8-1348 k^7\\
   &+4091 k^6-7768k^5+8451 k^4-3560 k^3-1612 k^2+1872 k-432\big),\\
Q_5(j)=& 2 j^{20}-795 j^{19}+40760 j^{18}-1036641 j^{17}+16752826
   j^{16}-191239786 j^{15}+1632641752 j^{14}\\
   &-10786299042 j^{13}+56334695030
   j^{12}-235648109263 j^{11}+795075807544 j^{10}\\
   &-2168602473357
   j^9+4771126881598 j^8-8409573468828 j^7+11731291260824
   j^6\\
   &-12705852943232 j^5+10375981856560 j^4-6104512549760
   j^3+2399836168064 j^2\\
   &-547585520256 j+51445094400,\\
Q_6(i)=&(i-5) (i-1) i (i+1) \big(16 i^{33}-7192 i^{32}+673840
   i^{31}-33108234 i^{30}+1069628658 i^{29}\\
   &+\dots+162245083333715039232 i-11706508031797555200\big) \big(16
   i^{33}-6664 i^{32}+452144 i^{31}\\
   &-15699130 i^{30}+\dots+6071537402380800 i^2-670382971978752
   i+32623028121600\big).
\end{align*}}%
\end{example}

\begin{example}\label{Exp:35Order}
The solution of the recurrence for the $C_F^3$-contribution to the unpolarized
3-loop Wilson coefficient for deeply inelastic scattering,
$C_{2,q,C_F^3}^{(3)}(N)$, constituted the hardest problem to solve. We obtained a
recurrence of order~35. Then our solver ran 25~hours and used 3\,GB of memory to derive the
35 linearly independent solutions. In total, we needed only 478 instead
of $\sum_{i=0}^{34}i=595$
summation quantifiers in order to represent those solutions. This is possible due to
the Improvement~1. For each of the summands around 20\,MB of memory were used.
In particular, in the summands the denominators have irreducible factors up to
degree~1000; the integer coefficients of the polynomials were up to 700 decimal digits long.
\end{example}

\subsection{Simplification of d'Alembertian solutions}

We consider the following problem: \textit{Given} indefinite nested sum and product expressions, e.g.,  expressions of the form~\eqref{Equ:dAlembertSol}, \textit{find} an alternative sum representation with the following properties:
\begin{enumerate}
\item All the involved sums are algebraically independent with each other.
\item The nested depth of the sum expressions is minimal.
\item In the summands the degree of the denominators is minimal.
\item The sums should be tuned in such a way that algorithms can perform this simplification as efficiently as possible.
\end{enumerate}
In principal, this problem can be solved with Karr's summation
algorithm~\cite{Karr:81} based on $\Pi\Sigma$-difference fields, if one knows explicitly the sum elements in which, e.g., the expression~\eqref{Equ:dAlembertSol} should be expressed.
For small examples such optimal sums with properties 1--3 from above might be guessed. In particular, if one has additional knowledge about the objects under consideration, a good sum representation might be known a priory.
But if such additional knowledge is not available, Karr's algorithm is not applicable.

In order to overcome this restriction, the fourth named author has refined Karr's $\Pi\Sigma$-theory for symbolic summation~\cite{Schneider:08e,Schneider:07d}. As a consequence, we can determine completely automatically such sum representations with the properties $1$--$4$ from above; see~\cite{Schneider:T07f,Schneider:08d}.

\begin{example}
With \SigmaP\ we find the depth-optimal representation
\begin{align*}
-\frac{4 \left(N^2+N+2\right)}{3 (N-1) N (N+1)}\left(\sum_{i=1}^N\frac{1}{i}\right)^2
+\frac{8
   \left(8 N^3+13 N^2+27 N+16\right)}{9 (N-1) N
   (N+1)^2}\sum_{i=1}^N\frac{1}{i}\\
   -\frac{8 \left(4 N^4+4 N^3+23 N^2+25 N+8\right)}{9 (N-1)N (N+1)^3}-\frac{4 \left(N^2+N+2\right)}{3 (N-1) N (N+1)}\sum_{i=1}^N\frac{1}{i^2}
\end{align*}
of~\eqref{Equ:Solution1} where the sums are given in~\eqref{Equ:DoubleSumSol}. We can read off the harmonic sum representation
\begin{align*}
&-\frac{4 \left(N^2+N+2\right)}{3(N-1)N(N+1)}S_1(N)^2
+\frac{8
   \left(8 N^3+13 N^2+27 N+16\right)}{9 (N-1) N
   (N+1)^2}S_1(N)\\
&-\frac{8 \left(4 N^4+4 N^3+23 N^2+25 N+8\right)}{9 (N-1)N (N+1)^3}-\frac{4 \left(N^2+N+2\right)}{3 (N-1) N (N+1)}S_2(N).
\end{align*}
\end{example}

\begin{example}
The sum expression for $P_{NS,2,C_A C_FN_F}^{-}$ containing in
particular the 7-nested sum \eqref{Equ:Solution2} can be simplified with
\SigmaP\ to the depth-optimal representation
\allowdisplaybreaks[3]
\begin{align*}
P&_{NS,2,C_A C_FN_F}^{-}=-\frac{2 \left(1086 N^7+3258 N^6+2129 N^5-288
N^4-67
   N^3-206 N^2-156 N+144\right)}{27 N^4 (N+1)^3}\\
&\frac{32 \left(8 N^4+33 N^3+53 N^2+25 N+3\right)}{9 N
   (N+1)^4}(-1)^N
   +\frac{16}{3}
   \sum_{i=1}^N\frac{1}{i^4}+\frac{32}{3}
   \sum_{i=1}^N\frac{(-1)^i}{i^4}\\
&   -\frac{16\left(10 N^2+10 N+3\right)}{9
N(N+1)}\sum_{i=1}^N\frac{(-1)^i}{i^3}+\frac{1336}{27}
   \sum_{i=1}^N\frac{1}{i^2}-\frac{64
   \left(8 N^2+8 N+3\right)}{9 N(N+1)}\sum_{i=1}^N\frac{(-1)^i}{i}\\
&   +\frac{16 \left(4
   N^6+88 N^5+314 N^4+412 N^3+201 N^2+16 N-12\right)}{9 N^2
   (N+1)^2 (N+2)^2}\sum_{i=1}^N\frac{(-1)^i}{i^2}\\
&   +
   \left(-\frac{8 \left(14 N^2+14 N+3\right)}{3 N
(N+1)}-\frac{16}{3}\sum_{i=1}^N\frac{1}{i}\right)\sum_{i=1}^N\frac{1}{i^3}+\frac{64}{3}
   \sum_{i=1}^N\frac{\sum_{j=1}^i\frac{1}{j^3}}{i}
   +32\sum_{i=1}^N\frac{\sum_{j=1}^i\frac{(-1)^j}{j^2}}{(i+2)^2}\\
&   -\frac{32 \left(22 N^2+22
   N-3\right)}{9 N
(N+1)}\sum_{i=1}^N\frac{\sum_{j=1}^i\frac{(-1)^j}{j^2}}{i+2}+\left(\sum_{i=1}^N\frac{1}{i}\right)\Bigg(\frac{32\left(2
N^2+4 N+1\right)}{3
   (N+1)^3}(-1)^N\\
&   -\frac{4 \left(65 N^6+195 N^5+195 N^4+137 N^3+36 N^2+36
   N+18\right)}{27 N^3 (N+1)^3}+\frac{32}{3}\sum_{i=1}^N
   \frac{(-1)^i}{i^3}+\frac{128}{3}
   \sum_{i=1}^N\frac{(-1)^i}{i}\\
&   +\frac{32
   \left(2 N^3+2 N^2-3 N-2\right)}{3 N
(N+1)(N+2)}\sum_{i=1}^N\frac{(-1)^i}{i^2}-\frac{64}{3}
   \sum_{i=1}^N\frac{\sum_{j=1}^i\frac{(-1)^j}{j^2}}{i+2}\Bigg)\\
&   -\frac{256}{3}
   \sum_{i=1}^N\frac{(-1)^i
   \sum_{j=1}^i\frac{1}{j}}{i}
   +\frac{128}{3}
   \sum_{i=1}^N\frac{\left(\sum_{j=1}^i\frac{(-1)^j}{j^2}\right)\left(
   \sum_{i=1}^N\frac{1}{j}\right)}{i+2}.
\end{align*}
Finally, we use J.~Ablinger's \textsf{HarmonicSums}
package~\cite{Ablinger:09}~\footnote{The package refers to algorithms and methods
from \cite{summer,ALGEBRA,STRUCT5,STRUCT6,VR,HOFM}.},
which transforms this expression to the
harmonic sum notation:
{\small
\begin{align*}
&\frac{64 (-1)^N (4 N+1)}{9 (N+1)^4}-\frac{2\left(270
   N^7+810 N^6-463 N^5-1392 N^4-211 N^3-206 N^2-156
   N+144\right)}{27 N^4 (N+1)^3}\\
&+\frac{64}{3}
   S_{-4}(N)+S_{-3}(N) \left(\frac{32}{3} S_{1}(N)-\frac{16
   \left(10 N^2+10 N+3\right)}{9 N
   (N+1)}\right)+\frac{32
   \left(10 N^2+10 N-3\right)}{9 N (N+1)}S_{-2,1}(N)\\
&+S_{-2}(N)
   \left(\frac{16 \left(16 N^2+10 N-3\right)}{9 N^2
   (N+1)^2}-\frac{320}{9} S_{1}(N)+\frac{64}{3}
   S_{2}(N)\right)-\frac{8 \left(14 N^2+14 N+3\right)}{3 N (N+1)}S_{3}(N)\\
&+S_{1}(N)
   \Big(\frac{-4 \left(209 N^6+627 N^5+627 N^4+281 N^3+36
   N^2+36 N+18\right)}{27 N^3 (N+1)^3}+16
   S_{3}(N)+\frac{80}{3} S_{4}(N)\\
&+\frac{1336}{27} S_{2}(N)+\frac{64}{3} S_{-2,1}(N)-\frac{32 (-1)^N}{3
   (N+1)^3}\Big)-\frac{32}{3} S_{2,-2}(N)
   -\frac{64}{3}
   S_{3,1}(N)-\frac{128}{3} S_{-2,1,1}(N).
\end{align*}
\normalsize}%
We emphasize that the harmonic sums in this expression are algebraically
independent. The algebraic independence could be accomplished with the
\SigmaP\ package; out of convenience and efficiency we used the
\textsf{HarmonicSums} package which contains among various other
features the harmonic sum relations of~\cite{ALGEBRA}.
\allowdisplaybreaks[3]
\end{example}

\begin{example}
The derived sum expression of $C_{2,q,C_F^3}^{(3)}$ from
Example~\ref{Exp:35Order} contains sums of the
form~\eqref{Equ:dAlembertSol} with depth $k=35$. In around four days and 20~hours
this expression could be simplified to an expression in terms of
65~sums that satisfy the properties 1--3 from above. Among them
there are 47~sums with depth two; typical examples are
\begin{equation}\label{Equ:2Nested}
\sum _{k=1}^N \frac{\displaystyle\left(\sum _{j=1}^k
\frac{(-1)^j}{j^2}\right)
   \left(\sum _{j=1}^k \frac{1}{j}\right)^3}{k}\quad\text{ and }\quad
\sum _{k=1}^N
   \frac{\displaystyle\left(\sum _{j=1}^k \frac{1}{j^2}\right)
\left(\sum _{j=1}^k
   \frac{(-1)^j}{j^2}\right) \sum _{j=1}^k \frac{1}{j}}{k}.
\end{equation}
Only one sum of nested depth three has been used, namely
$$\sum _{k=1}^N
   \frac{\displaystyle\left(\sum _{j=1}^k \frac{(-1)^j}{j^2}\right) \sum
_{j=1}^k
   \frac{\displaystyle\sum _{i=1}^j \frac{1}{i^2}}{j}}{k}.$$
We emphasize that these sums are constructed in such a way that the
difference field algorithms~\cite{Schneider:08e} work most efficiently:
The less nested the sums are, the more efficient our algorithms work.
E.g., if we switch to harmonic sum notation with Ablinger's
\textsf{HarmonicSum} package, the first sum in~\eqref{Equ:2Nested} can
be rewritten as
\begin{align*}
&S_{-3}(N) S_{3}(N)-S_{-2,1}(N) S_{3}(N)+
  S_{-2}(N)\big(\tfrac{1}{4} S_{1}(N)^4-\tfrac{3}{4} S_{2}(N)^2-\tfrac{3}{2} \
S_{4}(N)\big)\\
&+S_{4,-2}(N)-3 S_{-3,1,2}(N)-3 S_{-3,2,1}(N)+3
S_{-2}(N) S_{2,1,1}(N)-S_{3,1,-2}(N)\\
&+6 S_{-3,1,1,1}(N)+3S_{-2,1,1,2}(N)+3 S_{-2,1,2,1}(N)-6 S_{-2,1,1,1,1}(N);
\end{align*}
the involved sums have nested depths up to five. With such
representations the algorithms in \SigmaP\ work much slower, or might
even fail for our specific input due to time and memory limitations.
\end{example}
%%%%%%%%%%%%%%%%%%%%%%%%%%%%%%%%%%%%%%%%%%%%%%%%%%%%%%%%%%%%%%%%%%%%%%% %
%           Example
%%%%%%%%%%%%%%%%%%%%%%%%%%%%%%%%%%%%%%%%%%%%%%%%%%%%%%%%%%%%%%%%%%%%%%%
\section{\boldmath 3-Loop Anomalous Dimensions and Wilson Coefficients}
\label{sec:4}
%%%%%%%%%%%%%%%%%%%%%%%%%%%%%%%%%%%%%%%%%%%%%%%%%%%%%%%%%%%%%%%%%%%%%%%

\vspace{1mm}
\noindent
In the following we apply the method described in the previous section to
unfold all the unpolarized QCD anomalous dimensions and Wilson coefficients to
3-loop order from a series of Mellin moments. This sequence
is calculated using the relations given in \cite{MVV1,MVV2,MVV3} for the
different quantities per color factor and factors given by $\zeta$-values.
We will need rather high Mellin moments~$N$. The corresponding harmonic sums
cannot be calculated by {\tt summer} \cite{summer} directly, but have to be
evaluated recursively,
%------------------------------------------------------------------------------
\begin{eqnarray}
S_{a_1,a_2,...a_k}(N+1) =  \frac{{\rm sign}(a_1)^{N+1}}{(N+1)^{|a_1|}}
S_{a_2,...a_k}(N+1) +
S_{a_1,a_2,...a_k}(N)~.
\end{eqnarray}
%------------------------------------------------------------------------------
We used a \textsf{Maple} code for this. The highest moment to be calculated is
$N=5114$ for the
$C_F^3$--contribution to the 3-loop Wilson coefficient $C_{2,q}$. Its recursive
computation requires roughly 3~GB of memory and 270 min computational time on 
a 2 GHz processor. It is given by
a fraction with 13888 numerator and 13881 denominator digits. The set of
moments has a size of 69~MB. The determination of most of the other inputs
sets requires far less resources.

In Tables~1--3 we summarize the run parameters for the individual color- and
$\zeta$--contributions to the splitting functions
and in Tables~4--8 to the Wilson coefficients in unpolarized deeply inelastic
scattering up to 3--loop order. We specify the number of moments needed on
input and the order,
degree, and length of the recurrence derived. For the solution we compare
the number of harmonic sums in Refs.~\cite{MVV1,MVV2,MVV3} and in the present
calculation. The computation times needed to establish and to solve the
recurrences are also given.

To give some example for the rise of complexity for different orders in the 
coupling constant, we compare the $C_A^k$ contributions to 
$P_{gg,C_A^{k+1}}^{(k)}$.
In case of the anomalous dimensions the largest amount of moments needed is
$n = 19$ for $P_{gg,C_A}^{(0)}(n)$, $n = 181$ for $P_{gg,C_A^2}^{(1)}(n)$,
and $n = 1393$ for $P_{gg,C_A^3}^{(2)}(n)$. The order and degree of the
recurrences found are exactly, resp. nearly, the same for $P_{NS}^{(k),
\pm}(n)$. For the non-singlet anomalous dimensions and the singlet anomalous
dimensions and $P_{gq,gg}^{(k)}(n)$ order and degree of the difference equation
are larger than in case of $P_{qg}^{(k)}(n)$. The total computation time needed for
all anomalous dimensions amount to less than 18~h. The largest number of harmonic
sums contributing is 26. There are significant reductions in their number
comparing to the representation given in the attachment to
\cite{MVV1,MVV2}.\footnote{Here, the linear representation given in the text
has been reduced already, following an idea of one of the present authors.}
It amounts to a factor of two or larger, except in case of the very small
recurrences. In the non-singlet case $P_{NS,C_F^3}^{(k), \pm}(n)$
the number reduces from 68 to 26. A large reduction is obtained for
$P_{gg,C_A^3}^{(k), \pm}(n)$ from 130 to 21 harmonic sums.

For the Wilson coefficients $C_{2,q, C_F^3}^{(3)}(n), C_{2,q, C_F^2 C_A}^{(3)}(N)$
and $C_{2,q, C_F C_A^2}^{(3)}(n)$ four weeks of computation time is needed
in each case requiring $\leq 10$Gb on a 2~GHz processor.
The number of necessary harmonic sums is 60, reducing from 290
in \cite{MVV3} for $C_{2,q, C_F^2 C_A}^{(3)}$. This is the number of all
harmonic sums not containing the index $\{-1\}$ up to weight $w=6$ after
algebraic reduction, cf.\ \cite{STRUCT6}.

If one compares the number of harmonic sums obtained in the present
calculation after the algebraic reduction yields groups characterized by
clusters of 58-60, 26-29, 11-15 and cases with a number of sums below 10, up 
to very few exceptions.  As this
pattern is the same for quite different quantities, it may be related rather
to the topology, but the color- or field-structure of the respective diagrams.
This pattern is not seen counting the harmonic sums in the representation of
Ref.~\cite{MVV1,MVV2,MVV3}.

In case of the smaller recurrences the time needed for their derivation is
usually shorter than that for its solution. Conversely, for the larger recurrences
the time required to establish them and the solution time behave roughly like 4(3):1.
The
total computation time amounted to 110.3 CPU days. Concerning the size of the
different problems to be dealt with a naive fivefold parallelization was possible.
Here we did not yet consider parallelization w.r.t.\ the number of primes $N_p$
chosen, which would significantly reduce the computational time,
of the $C_F^3$ term of $C_{2,q}^{(3)}$,
with $N_p = 140$, discussed above and for other comparably large contributions.

In course of solving the recurrences we reduce the harmonic sums appearing
algebraically, \cite{ALGEBRA}, and can express all results in terms of the
following harmonics sums:
\begin{alignat*}{1}
& S_{1} \\
& S_{2}, S_{-2}\\
& S_{3}, S_{-3}\\
& S_{2,1}, S_{1,-2}\\
& S_{4}, S_{-4}\\
& S_{3,1}, S_{-3,1}, S_{2,-2}\\
& S_{2,1,1}, S_{-2,1,1}\\
& S_{5}, S_{-5}\\
& S_{4,1}, S_{-4,1}, S_{3,-2}, S_{3,2}, S_{-3,2}, S_{-3,-2}\\
& S_{3,1,1}, S_{-3,1,1}, S_{2,2,1}, S_{-2,1,-2}, S_{2,1,-2}, S_{-2,2,1}\\
& S_{2,1,1,1}, S_{-2,1,1,1}\\
& S_{6}, S_{-6}\\
& S_{5,1}, S_{-5,1}, S_{4,2}, S_{4,-2}, S_{-4,2}, S_{-4,-2}, S_{-3,3} \\
& S_{4,1,1}, S_{-4,1,1}, S_{3,2,1}, S_{2,3,1}, S_{-3,2,1}, S_{-3,1,2},
 S_{-2,3,1}, S_{3,1,-2}, S_{-3,1,-2}, S_{-3,-2,1}, S_{-2,2,2}, S_{2,-2,-2}\\
& S_{3,1,1,1}, S_{-3,1,1,1}, S_{2,2,1,1}, S_{-2,-2,1,1}, S_{2,-2,1,1}, S_{-2,2,1,1},
  S_{-2,1,1,2}\\
& S_{2,1,1,1,1}, S_{-2,1,1,1,1}~.
\end{alignat*}
%-------------------------------------------------------------------------------
The 3-loop Wilson coefficients require the complete set of possible functions
up to $w = 6$. This representation can be further reduced using the
structural relations \cite{STRUCT5,STRUCT6} to:
\begin{alignat*}{1}
& S_{1}\\
& S_{2,1}, S_{-2,1}\\
& S_{-3,1}\\
& S_{2,1,1}, S_{-2,1,1}\\
& S_{4,1}, S_{-4,1}\\
& S_{3,1,1}, S_{-3,1,1}, S_{2,2,1}, S_{-2,1,-2}, S_{2,1,-2}, S_{-2,2,1}\\
& S_{2,1,1,1}, S_{-2,1,1,1}\\
& S_{-5,1} \\
& S_{4,1,1}, S_{-4,1,1}, S_{3,2,1}, S_{2,3,1}, S_{-3,2,1}, S_{-3,1,2},
 S_{-2,3,1}, S_{3,1,-2}, S_{-3,1,-2}, S_{-3,-2,1}, S_{-2,2,2}, S_{2,-2,-2}\\
& S_{3,1,1,1}, S_{-3,1,1,1}, S_{2,2,1,1}, S_{-2,-2,1,1}, S_{2,-2,1,1}, S_{-2,2,1,1},
  S_{-2,1,1,2}\\
& S_{2,1,1,1,1}, S_{-2,1,1,1,1}~.
\end{alignat*}
%-------------------------------------------------------------------------------
In \cite{STRUCT5,STRUCT6} we applied a slightly different basis referring to
$S_{-2,2,-2}$ instead of $S_{2,-2,-2}$ and to $S_{2,-3,1}$ instead of
$S_{-3,1,2}$, which is algebraically equivalent. These 38 functions can be
represented by 35 basic Mellin transforms.

The ab-initio calculation of moments for
the quantities considered in the present paper can be performed by codes like
\textsf{mincer} and
\textsf{MATAD} \cite{MINMAT} available for physics calculations. Both the
computational time and memory requests rise drastically going to higher values
of $N$. In case of \textsf{mincer} both parameters increase by a factor of
$\sim 5$ enlarging $N \rightarrow N+2$. Comparable, but slightly larger factors
are obtained for \textsf{MATAD}.
In the well--known leading order case,
enough moments may be provided for our procedure. Already for some color
projections of the next-to-leading order corrections, this is no longer the
case,~\cite{VERM}, since around 150 initial values are needed. For the 3-loop
anomalous dimensions and Wilson coefficients $N=16$ can be reached with computation
times of the order of 0.5--1 CPU year, cf.~\cite{MOM1}. The codes
\cite{MINMAT} still may be improved. However, the power--growth going to
higher moments will basically remain due to the algorithms used. The method 
presented
in this paper can therefore not be applied to {\sf whole} color-factor 
contributions for
the anomalous dimensions and Wilson coefficients at the 3--loop level. They may,
however, be useful in solving medium-size problems. In view of constructing general
methods suitable to evaluate single scale quantities,
methods to evaluate the fixed
moments for these quantities at far lower expenses have to be developed.

To illustrate the results of the present calculation, the non-singlet anomalous dimensions
to $O(a_s^3)$ are given as an example in the appendix. The relations for all unpolarized
anomalous dimensions and Wilson coefficients, separated according to the corresponding
color- and $\zeta$-value terms, are attached to this paper in \FORMP- and
\mathP\ files. The \FORMP-codes provide a check of our relations with the moments
calculated in Ref.~\cite{LRV,MOM}.

%%%%%%%%%%%%%%%%%%%%%%%%%%%%%%%%%%%%%%%%%%%%%%%%%%%%%%%%%%%%%%%%%%%%%%% %
%           Conclusions
%%%%%%%%%%%%%%%%%%%%%%%%%%%%%%%%%%%%%%%%%%%%%%%%%%%%%%%%%%%%%%%%%%%%%%%
\section{Conclusions}
\label{sec:5}
%%%%%%%%%%%%%%%%%%%%%%%%%%%%%%%%%%%%%%%%%%%%%%%%%%%%%%%%%%%%%%%%%%%%%%%

\vspace{1mm}
\noindent
We established a general algorithm to calculate the exact expression
for single scale quantities from a finite, suitably large number of moments,
which are zero scale quantities. The latter ones are much more easily calculable
than single scale quantities. We applied the method to the anomalous
dimensions and Wilson coefficients up to 3-loop order. Hereby we compactified
their representation exploiting all algebraic relations between the harmonic
sums. The 3-loop Wilson coefficients require the whole set of basic harmonic
sums in the sub-algebra spanned by the index set to $w=6$ without $i=-1$.
A further compactification can be obtained using the structural relations
between the harmonic sums. After algebraic reduction the number of the
harmonic sums contributing clusters in several classes mainly determined by
the topology of the graphs and widely independent of
the color- and field structure of the respective contributions. The
CPU time for the whole problem amounted to about four months using 2~GHz processors 
and $\lsim$ 10 GB of memory were needed. The problem can be
naively parallelized fivefold. The real computational time needed to establish
the recurrences can be shortened further running Chinese remaindering in
parallel.

To solve 3-loop problems for whole color factor contributions is not 
possible at present, since the number
of required moments is too large for the methods available. Methods to evaluate
the fixed moments for these quantities to high order at far lower expenses
have still to be developed.

We established and solved the recurrences for all color resp. $\zeta$-projections 
at once, which
forms a rather voluminous problem. Yet we showed that rather large difference
equations [order 35; degree $\sim$ 1000], which occur for the most advanced
problems in Quantum Field Theory, can be reliably and fast established
and solved unconditionally.

\vspace{5mm}\noindent
\textbf{Acknowledgments.}
We would like to thank J.~Vermaseren for discussions.
This work was supported in part by DFG Sonderforschungsbereich Transregio~9,
Computergest\"utzte Theoretische Teilchenphysik, projects P19462-N18, P20162-N18, and
P20347-N18 of the Austrian~FWF, and Studienstiftung des Deutschen Volkes.

\afterpage{\clearpage} % force tables now.

%%%%%%%%%%%%%%%%%%%%%%%%%%%%%%%%%%%%%%%%%%%%%%%%%%%%%%%%%%%%%%%%%%%%%%%%%%%%%
%%%%%%%%%%%%%%%%%%%%  TABLE 1 %%%%%%%%%%%%%%%%%%%%%%%%%%%%%%%%%%%%%%%%%%%%%%%
\begin{table}[ht]
\caption{\sf Run parameters for the unfolding of the non-singlet
             anomalous dimensions}
%\label{table:TAB:NS}
{\footnotesize
\begin{center}
\renewcommand{\arraystretch}{1.3}
\begin{tabular}{||l|r|r|r|r|r|r|r||}
\hline \hline
\multicolumn{1}{||c|}{ } &
\multicolumn{1}{c|}{number of}  &
\multicolumn{1}{c|}{order of}   &
\multicolumn{1}{c|}{degree of}  &
\multicolumn{1}{c|}{total time}&
\multicolumn{1}{c|}{length of} &
\multicolumn{1}{c|}{number of} &
\multicolumn{1}{c||}{solution}
\\
\multicolumn{1}{||c|}{ } &
\multicolumn{1}{c|}{terms} &
\multicolumn{1}{c|}{recurrence}   &
\multicolumn{1}{c|}{recurrence}   &
\multicolumn{1}{c|}{[sec]}       &
\multicolumn{1}{c|}{recurrence}   &
\multicolumn{1}{c|}{harm. sums}   &
\multicolumn{1}{c||}{time [sec]}   \\
\multicolumn{1}{||c|}{ } &
\multicolumn{1}{c|}{needed} &
\multicolumn{1}{c|}{ }   &
\multicolumn{1}{c|}{ }   &
\multicolumn{1}{c|}{ }       &
\multicolumn{1}{c|}{[kbyte]}   &
\multicolumn{1}{c|}{a [b] }   &
\multicolumn{1}{c||}{ }   \\
\hline \hline
$P_{NS,0}$                         &   14&   2&   3&    0.05&0.087 &1 [1] &0.55  \\
\hline
$P_{NS,1,C_F^2}^{-}$               &  142&   5&  31&    3.32&4.666 &6 [10] & 7.45\\
$P_{NS,1,C_A C_F}^{-}$             &  109&   4&  24&    1.91&2.834 &6 [7] & 6.28\\
$P_{NS,1,C_F N_F}^{-}$             &   24&   2&   7&    0.13&0.271 &2 [2] &0.92\\
\hline
$P_{NS,1,C_F^2}^{+}$               &  142&   5&  31&    3.35&4.707 &6 [10] & 7.45\\
$P_{NS,1,C_A C_F}^{+}$             &  109&   4&  23&    1.88&2.703 &6 [7]  & 6.23\\
$P_{NS,1,C_F N_F}^{+}$             &   24&   2&   7&    0.09&0.271 &2 [2]  & 0.89\\
\hline
$P_{NS,2,C_F^3}^{-}$               & 1079&  16& 192& 3152.19&529.802 &26 [68] & 1194.41\\
$P_{NS,2,C_F^3 \zeta_3}^{-}$       &   48&   3&  11&    0.49&0.643 &1 [1] & 1.56\\
$P_{NS,2,C_A C_F^2}^{-}$           &  974&  15& 181& 1736.08&450.919 &26 [62] & 1194.41\\
$P_{NS,2,C_A C_F^2 \zeta_3}^{-}$   &   48&   3&  11&    0.53&0.643 &1 [1] & 1.53\\
$P_{NS,2,C_A^2 C_F}^{-}$           &  749&  12& 147& 1004.12&242.892 &26 [62] & 1100.88 \\
$P_{NS,2,C_A^2 C_F \zeta_3}^{-}$   &   48&   3&  11&    0.56&0.643 &1 [1] & 1.56\\
$P_{NS,2,C_F N_F^2}^{-}$           &   39&   2&  11&    0.31&0.369 &3 [3] & 1.20\\
$P_{NS,2,C_F^2 N_F}^{-}$           &  377&   8&  68&   76.34&33.946 &11 [24] & 72.22\\
$P_{NS,2,C_F^2 N_F \zeta_3}^{-}$   &   14&   2&   3&    0.12&0.101 &1 [1]& 0.53\\
$P_{NS,2,C_A C_F N_F}^{-}$         &  356&   7&  62&   65.25&23.830 &11 [20] & 52.67\\
$P_{NS,2,C_A C_F N_ F \zeta_3}^{-}$&   14&   2&   3&    0.12&0.101 &1 [1] &0.55 \\
\hline
$P_{NS,2,C_F^3}^{+}$               & 1079&  16& 192& 4713.27&527.094 &26[68] & 1165.22\\
$P_{NS,2,C_F^3 \zeta_3}^{+}$       &   48&   3&  11&    0.55&0.643 &1[1] &1.562 \\
$P_{NS,2,C_A C_F^2}^{+}$           &  974&  15& 178& 1715.03&442.031 &26[62] &889.047 \\
$P_{NS,2,C_A C_F^2 \zeta_3}^{+}$   &   48&   3&  11&    0.61&0.643 &1[1] &1.531 \\
$P_{NS,2,C_A^2 C_F}^{+}$           &  749&  12& 146&  991.22&240.325 &26[50] & 516.812\\
$P_{NS,2,C_A^2 C_F \zeta_3}^{+}$   &   48&   3&  11&    0.61&0.643 &1[1] &1.593 \\
$P_{NS,2,C_F^2 N_F}^{+}$           &  377&   8&  69&  111.38&33.872 &11[24] &71.235 \\
$P_{NS,2,C_F^2 N_F \zeta_3}^{+}$   &   14&   2&   3&    0.15&0.101 &1[1] &0.531 \\
$P_{NS,2,C_A C_F N_F}^{+}$         &  307&   7&  61&   48.62&23.808 &11[24] &71.235 \\
$P_{NS,2,C_A C_F N_ F \zeta_3}^{+}$&   14&   2&   3&    0.15&0.101 &1[1] &0.547 \\
$P_{NS,2,C_F N_F^2}^{+}$           &   39&   2&  11&    0.40&0.369 &3[3] &1.172 \\
\hline
$P_{NS,2,N_F d_{abc}}^{-}$         &   459&  7&   87& 239.62&0.369 &5 [20] &
32.5\\
\hline
\hline
\end{tabular}
\renewcommand{\arraystretch}{1.0}
\end{center}
}
\end{table}
%%%%%%%%%%%%%%%%%%%%%%%%%%%%%%%%%%%%%%%%%%%%%%%%%%%%%%%%%%%%%%%%%%%%%%%%%%%%%

%%%%%%%%%%%%%%%%%%%%%%%%%%%%%%%%%%%%%%%%%%%%%%%%%%%%%%%%%%%%%%%%%%%%%%%%%%%%%
%%%%%%%%%%%%%%%%%%%%  TABLE 2 %%%%%%%%%%%%%%%%%%%%%%%%%%%%%%%%%%%%%%%%%%%%%%%
\begin{table}[ht]
\caption{\sf Run parameters for the unfolding of the singlet
             anomalous dimensions}
%\label{table:TAB:NS}
{\footnotesize
\begin{center}
\renewcommand{\arraystretch}{1.3}
\begin{tabular}{||l|r|r|r|r|r|r|r||}
\hline \hline
\multicolumn{1}{||c|}{ }        &
\multicolumn{1}{c|}{number of}  &
\multicolumn{1}{c|}{order of}   &
\multicolumn{1}{c|}{degree of}  &
\multicolumn{1}{c|}{total time} &
\multicolumn{1}{c|}{length of}  &
\multicolumn{1}{c|}{number of}  &
\multicolumn{1}{c||}{solution}
\\
\multicolumn{1}{||c|}{ }          &
\multicolumn{1}{c|}{terms} &
\multicolumn{1}{c|}{recurrence}   &
\multicolumn{1}{c|}{recurrence}   &
\multicolumn{1}{c|}{[sec]}        &
\multicolumn{1}{c|}{recurrence}   &
\multicolumn{1}{c|}{harm. sums}   &
\multicolumn{1}{c||}{time [sec]}   \\
\multicolumn{1}{||c|}{ } &
\multicolumn{1}{c|}{needed} &
\multicolumn{1}{c|}{ }   &
\multicolumn{1}{c|}{ }   &
\multicolumn{1}{c|}{ }       &
\multicolumn{1}{c|}{[kbyte]}   &
\multicolumn{1}{c|}{a [b] }   &
\multicolumn{1}{c||}{ }   \\
\hline \hline
 $P^{PS}_{1,N_F C_F}$      &   24 &  1 &   8 &     0.19 &0.204 &0[0] &0.244 \\
\hline
 $P^{PS}_{2,N_F^2 C_F}$    &  109 &  3 &  26 &     6.32 &1.891 &2[8] &2.812 \\
 $P^{PS}_{2,N_F C_A C_F}$    &  566 &  9 & 115 &   425.44 &100.414 &7 [40] & 111.52\\
 $P^{PS}_{2,N_F C_A C_F \zeta_3}$  &   19 &  1 &   6 &     0.42 &0.117 &0[0] &0.204\\
 $P^{PS}_{2,N_F C_F^2}$      &  237 &  5 &  50 &    32.75  &0.117 &4[24] &14.601\\
 $P^{PS}_{2,N_F C_F^2 \zeta_3}$   &   19 &  1 &   6 &     0.41  &12.163 &0[0] &0.200\\
\hline
 $P_{qg,0}$                     &   11 &  1 &   3 &     0.02  &0.061 &0 [0] &0.16\\
\hline
 $P_{qg,1,N_F C_A}$             &  120 &  4 &  29 &     3.31  &3.769 &3[8] &4.872\\
 $P_{qg,1,N_F C_F}$             &   63 &  3 &  16 &     0.68  &0.951 &2[9] & 2.008\\
\hline
 $P_{qg,2,N_F^2 C_A}$            &  239 &  6 &  54 &    45.30  &17.403 &6[24] & 21.993\\
 $P_{qg,2,N_F^2 C_F}$            &  194 &  5 &  41 &    27.75  &7.911 &3[15] &8.021\\
 $P_{qg,2,N_F C_A^2}$            & 1088 & 15 & 201 &  3321.46  &557.535 &13 [88] &848.85\\
 $P_{qg,2,N_F C_A^2 \zeta_3}$    &   39 &  2 &  11 &     0.86  &0.408 &1[3] &0.932\\
 $P_{qg,2,N_F C_A C_F}$         & 1049 & 15 & 194 &  3963.62   &552.100 &12 [84] & 714.45\\
 $P_{qg,2,N_F C_A C_F \zeta_3}$  &   39 &  2 &  11 &     1.02  &0.409 &1 [3] & 0.93\\
 $P_{qg,2,N_F C_F^2}$            &  849 & 12 & 143 &  1337.36  &261.804 &13 [66] &387.6\\
 $P_{qg,2,N_F C_F^2 \zeta_3}$    &   17 &  1 &   5 &     0.29  &0.093 &0 [0] & 0.15\\
\hline
 $P_{gq,0}$                     &   11 &  1 &   3 &     0.03  & 0.062&0 [0] & 0.15\\
\hline
 $P_{gq,1,C_F^2}$               &   63 &  3 &  16 &     0.72  &0.869 &2[6] &1.924 \\
 $P_{gq,1,C_F C_A}$             &  125 &  4 &  31 &     4.55  &4.059 &3[12] &5.068 \\
 $P_{gq,1,N_F C_F}$             &   24 &  2 &   6 &     0.18  &0.192 &1[3] &0.588 \\
\hline
 $P_{gq,2,C_F^3}$               &  703 & 11 & 114 &   927.82  &162.320 &13[59] &245.53 \\
 $P_{gq,2,C_F^3 \zeta_3}$       &   35 &  2 &   9 &     0.79  &0.281 &1[3] &0.776\\
 $P_{gq,2,C_A^2 C_F}$           & 1088 & 15 & 203 &  3327.32  &633.346 &12 [93] & 830.2\\
 $P_{gq,2,C_A^2 C_F \zeta_3}$   &   35 &  2 &   9 &     0.81  &0.281 &1[3] &0.776 \\
 $P_{gq,2,C_A C_F^2}$           & 1087 & 15 & 193 &  3184.13  &528.827 &14 [75] & 853.31\\
 $P_{gq,2,C_A C_F^2 \zeta_3}$   &   35 &  2 &   9 &     0.86  &0.281 &1[3] & 0.776\\
 $P_{gq,2,N_F C_F^2}$           &  339 &  7 &  69 &   106.93  &30.626 &5[25] &33.586 \\
 $P_{gq,2,N_F C_F^2 \zeta_3}$   &   11 &  1 &   3 &     0.24  &0.062 &0[0] & 0.152 \\
 $P_{gq,2,N_F C_A C_F}$         &  1087&  15 &  194 & 3201.52   &58.943 &17[87] &714.51 \\
 $P_{gq,2,N_F C_A C_F \zeta_3}$ &   11 &  1 &   3 &     0.21  &0.062 &0[0]&0.156 \\
 $P_{gq,2,N_F^2 C_F}$           &   41 &  3 &   9 &     1.04  &0.445 &2[6] &1.216 \\
\hline
 $P_{gg,0}$                     &   19 &  2 &   5 &     0.04 & 0.166&1 [1] & 0.65\\
\hline
 $P_{gg,1, C_A^2}$              &  181 &  5 &  45 &    12.07 & 9.053&6 [17] & 11.62\\
 $P_{gg,1, N_F C_A}$            &   29 &  2 &   9 &     0.23 & 0.395&1[1] & 0.856\\
 $P_{gg,1, N_F C_F}$            &   31 &  1 &  11 &     0.23 & 0.228&0[0] & 0.240\\
\hline
\hline
\end{tabular}
\renewcommand{\arraystretch}{1.0}
\end{center}
}
\end{table}
%%%%%%%%%%%%%%%%%%%%%%%%%%%%%%%%%%%%%%%%%%%%%%%%%%%%%%%%%%%%%%%%%%%%%%%%%%%%%
%\afterpage{\clearpage} % force tables now.

%%%%%%%%%%%%%%%%%%%%%%%%%%%%%%%%%%%%%%%%%%%%%%%%%%%%%%%%%%%%%%%%%%%%%%%%%%%%%
%%%%%%%%%%%%%%%%%%%%  TABLE 3 %%%%%%%%%%%%%%%%%%%%%%%%%%%%%%%%%%%%%%%%%%%%%%%
\begin{table}[ht]
\caption{\sf Run parameters for the unfolding of the singlet
             anomalous dimensions (continued)}
%\label{table:TAB:NS}
{\footnotesize
\begin{center}
\renewcommand{\arraystretch}{1.3}
\begin{tabular}{||l|r|r|r|r|r|r|r||}
\hline \hline
\multicolumn{1}{||c|}{ }        &
\multicolumn{1}{c|}{number of}  &
\multicolumn{1}{c|}{order of}   &
\multicolumn{1}{c|}{degree of}  &
\multicolumn{1}{c|}{total time} &
\multicolumn{1}{c|}{length of}  &
\multicolumn{1}{c|}{number of}  &
\multicolumn{1}{c||}{solution}
\\
\multicolumn{1}{||c|}{ }          &
\multicolumn{1}{c|}{terms} &
\multicolumn{1}{c|}{recurrence}   &
\multicolumn{1}{c|}{recurrence}   &
\multicolumn{1}{c|}{[sec]}        &
\multicolumn{1}{c|}{recurrence}   &
\multicolumn{1}{c|}{harm. sums}   &
\multicolumn{1}{c||}{time [sec]}   \\
\multicolumn{1}{||c|}{ } &
\multicolumn{1}{c|}{needed} &
\multicolumn{1}{c|}{ }   &
\multicolumn{1}{c|}{ }   &
\multicolumn{1}{c|}{ }       &
\multicolumn{1}{c|}{[kbyte]}   &
\multicolumn{1}{c|}{a [b] }   &
\multicolumn{1}{c||}{ }   \\
\hline \hline
 $P_{gg,2,C_A^3}$               & 1393 & 16 & 277 & 12432.80 & 1087.615&21 [130] &2419.04\\
 $P_{gg,2,N_F C_F^2}$           &  439 &  8 &  88 &   237.82 & 57.291&7 [35]   &  55.52 \\
 $P_{gg,2,N_F C_F^2 \zeta_3}$   &   15 &  1 &   4 &     0.31 &0.073 &0[0] &0.156\\
 $P_{gg,2,N_F C_A^2}$           &  782 & 11 & 148 &  1638.62 &205.980 &6 [31] & 160.89\\
 $P_{gg,2,N_F C_A^2 \zeta_3}$   &   29 &  2 &   9 &     0.66 &0.308 &1[1] &0.796\\
 $P_{gg,2 N_F C_A C_F}$         &  749 & 10 & 127 &  1169.37 &146.921 &7 [40] &128.37\\
 $P_{gg,2 N_F C_A C_F \zeta_3}$ &   29 &  2 &   9 &     0.72 &0.305 &1[1] &0.828\\
 $P_{gg,2,N_F^2 C_A}$           &   55 &  2 &  17 &     4.53 &0.979 &1[4] &1.092\\
 $P_{gg,2,N_F^2 C_F}$           &  109 &  3 &  26 &     6.74 &2.483 &2[12] &2.668\\
\hline
\hline
\end{tabular}
\renewcommand{\arraystretch}{1.0}
\end{center}
}
\end{table}
%%%%%%%%%%%%%%%%%%%%%%%%%%%%%%%%%%%%%%%%%%%%%%%%%%%%%%%%%%%%%%%%%%%%%%%%%%%%%
%%%%%%%%%%%%%%%%%%%%%%%%%%%%%%%%%%%%%%%%%%%%%%%%%%%%%%%%%%%%%%%%%%%%%%%%%%%%%
%%%%%%%%%%%%%%%%%%%%  TABLE 4 %%%%%%%%%%%%%%%%%%%%%%%%%%%%%%%%%%%%%%%%%%%%%%%
\begin{table}[ht]
\caption{\sf Run parameters for the unfolding of the unpolarized pure-singlet
             Wilson Coefficients}
%\label{table:TAB:NS}
{\footnotesize
\begin{center}
\renewcommand{\arraystretch}{1.3}
\begin{tabular}{||l|r|r|r|r|r|r|r||}
\hline \hline
\multicolumn{1}{||c|}{ }        &
\multicolumn{1}{c|}{number of}  &
\multicolumn{1}{c|}{order of}   &
\multicolumn{1}{c|}{degree of}  &
\multicolumn{1}{c|}{total time} &
\multicolumn{1}{c|}{length of}  &
\multicolumn{1}{c|}{number of}  &
\multicolumn{1}{c||}{solution}
\\
\multicolumn{1}{||c|}{ }          &
\multicolumn{1}{c|}{terms} &
\multicolumn{1}{c|}{recurrence}   &
\multicolumn{1}{c|}{recurrence}   &
\multicolumn{1}{c|}{[sec]}        &
\multicolumn{1}{c|}{recurrence}   &
\multicolumn{1}{c|}{harm. sums}   &
\multicolumn{1}{c||}{time [sec]}   \\
\multicolumn{1}{||c|}{ } &
\multicolumn{1}{c|}{needed} &
\multicolumn{1}{c|}{ }   &
\multicolumn{1}{c|}{ }   &
\multicolumn{1}{c|}{ }       &
\multicolumn{1}{c|}{[kbyte]}   &
\multicolumn{1}{c|}{a [b] }   &
\multicolumn{1}{c||}{ }   \\
\hline \hline
 $C_{2,PS,C_F N_F}^{(2)}$              &  209 &  5 &  42 &    20.85 &8.422
&3[16]
&7.70 \\
\hline
 $C_{2,PS, C_F^2 N_F}^{(3)}$           & 1847 & 19 & 334 & 41001.00 &1989.043 &14 [122] & 2701.04 \\
 $C_{2,PS,C_F^2 N_F \zeta_3}^{(3)}$    &   65 &  2 &  20 &     0.69 &1.124 &1 [6]    &    0.92 \\
 $C_{2,PS,C_F^2 N_F \zeta_4}^{(3)}$    &   19 &  1 &   6 &     0.08 &0.117 &0 [0]    &    0.14\\
 $C_{2,PS, C_F C_A N_F}^{(3)}$         & 2023 & 20 & 368 & 54873.80 &2670.459 & 14 [126]& 4589.31 \\
 $C_{2,PS,C_F C_A N_F \zeta_3}^{(3)}$  &   71 &  2 &  21 &     0.82 &1.429 &  1 [6]   &    0.97 \\
 $C_{2,PS,C_F C_A N_F \zeta_4}^{(3)}$  &   19 &  1 &   6 &     0.08 &0.117 &  0 [0]   &    0.14 \\
 $C_{2,PS, C_F N_F^2}^{(3)}$           &  479 &  8 & 103 &   629.05 &75.646 &5 [34]    &   53.28 \\
 $C_{2,PS,C_F N_F^2 \zeta_3}^{(3)}$    &   19 &  1 &   6 &     0.09 &0.122 &0 [0]     &    0.14 \\
\hline
 $C_{L,PS,C_F N_F}^{(2)}$              &   41 &  2 &  11 &     0.20 &0.384 &1[4] & 0.88 \\
\hline
 $C_{L,PS,C_F^2 N_F}^{(3)}$            &  869 & 11 & 162 &  4411.20 &250.352 &8 [62] &163.17 \\
 $C_{L,PS,C_F^2 N_F \zeta_3}^{(3)}$    &   35 &  2 &  10 &     0.17 &0.406 &1 [5]  &0.63\\
 $C_{L,PS,C_F C_A N_F}^{(3)}$          &  840 & 11 & 153 &  2005.44 &231.837 & 8 [64] & 153.99\\
 $C_{L,PS,C_F C_A N_F \zeta_3}^{(3)}$  &   35 &  2 &  10 &     0.17 &0.403 &1 [5] &0.59 \\
 $C_{L,PS,C_F N_F^2}^{(3)}$            &  224 &  5 &  52 &    72.64 &12.440 &3 [13]  &9.87 \\
\hline
\hline
\end{tabular}
\renewcommand{\arraystretch}{1.0}
\end{center}
}
\end{table}
%%%%%%%%%%%%%%%%%%%%%%%%%%%%%%%%%%%%%%%%%%%%%%%%%%%%%%%%%%%%%%%%%%%%%%%%%%%%%
%%%%%%%%%%%%%%%%%%%%%%%%%%%%%%%%%%%%%%%%%%%%%%%%%%%%%%%%%%%%%%%%%%%%%%%%%%%%%
%%%%%%%%%%%%%%%%%%%%  TABLE 5 %%%%%%%%%%%%%%%%%%%%%%%%%%%%%%%%%%%%%%%%%%%%%%%
\begin{table}[ht]
\caption{\sf Run parameters for the unfolding of the unpolarized quarkonic
         Wilson Coefficients for the structure function $F_2(x,Q^2)$.}
%\label{table:TAB:NS}
{\footnotesize
\begin{center}
\renewcommand{\arraystretch}{1.3}
\begin{tabular}{||l|r|r|r|r|r|r|r||}
\hline \hline
\multicolumn{1}{||c|}{ }        &
\multicolumn{1}{c|}{number of}  &
\multicolumn{1}{c|}{order of}   &
\multicolumn{1}{c|}{degree of}  &
\multicolumn{1}{c|}{total time} &
\multicolumn{1}{c|}{length of}  &
\multicolumn{1}{c|}{number of}  &
\multicolumn{1}{c||}{solution}
\\
\multicolumn{1}{||c|}{ }          &
\multicolumn{1}{c|}{terms} &
\multicolumn{1}{c|}{recurrence}   &
\multicolumn{1}{c|}{recurrence}   &
\multicolumn{1}{c|}{[sec]}        &
\multicolumn{1}{c|}{recurrence}   &
\multicolumn{1}{c|}{harm. sums}   &
\multicolumn{1}{c||}{time [sec]}   \\
\multicolumn{1}{||c|}{ } &
\multicolumn{1}{c|}{needed} &
\multicolumn{1}{c|}{ }   &
\multicolumn{1}{c|}{ }   &
\multicolumn{1}{c|}{ }       &
\multicolumn{1}{c|}{[kbyte]}   &
\multicolumn{1}{c|}{a [b] }   &
\multicolumn{1}{c||}{ }   \\
\hline \hline
 $C_{2,q,C_F}^{(1)}$                  &   31 &  3 &   6 &     0.26 &0.429 & 2[3] &1.47 \\
\hline
 $C_{2,q,C_F^2}^{(2)}$                &  689 & 11 & 137 &  1134.10 &177.806
&13[39] &258.24 \\
 $C_{2,q,C_F^2 \zeta_3}^{(2)}$        &   15 &  2 &   3 &     0.27 &0.100
&1[1]
&0.54 \\
 $C_{2,q,C_A C_F}^{(2)}$              &  545 & 10 & 121 &   413.33 &127.893
&12[35]
&178.73 \\
 $C_{2,q,C_A C_F \zeta_3}$            &   15 &  2 &   3 &     0.27 &0.112
&1[1]
&0.55 \\
 $C_{2,q,N_F C_F}$                    &   71 &  4 &  16 &     2.68 &1.655
&4[10] &
3.95 \\
\hline
 $C_{2,q,C_F^3}^{(3)}$                & 5114 & 35 & 938 & 1.79 $\times 10^6$
&30394.173 &58[289] &509242 \\
 $C_{2,q,C_F^3 \zeta_3}^{(3)}$        &  284 &  8 &  64 &    31.02 &32.363 & 6 [18]& 27.60\\
 $C_{2,q,C_F^3 \zeta_4}^{(3)}$        &  65 &  3 &  11 &     2.62 &0.163 & 1 [1]& 1.47\\
 $C_{2,q,C_F^3 \zeta_5}^{(3)}$        &   19 &  2 &   5 &     0.08 &0.163 & 1 [1]& 0.47\\
 $C_{2,q,C_F^2 C_A}^{(3)}$            & 5059 & 35 & 930 &
1.69 $\times 10^6$ &30122.380 & 60
[290] &0.478 $\times  10^6$\\
 $C_{2,q,C_F^2 C_A \zeta_3}^{(3)}$    &  284 &  8 &  64 &    34.00 &33.400 & 7 [18]& 28.53\\
 $C_{2,q,C_F^2 C_A \zeta_4}^{(3)}$    &   48 &  3 &  11 &     0.32 &0.643 & 1[1] & 1.01\\
 $C_{2,q,C_F^2 C_A \zeta_5}^{(3)}$    &   19 &  2 &   5 &     0.08 &0.167 & 1 [1]& 0.42\\
 $C_{2,q,C_F C_A^2}^{(3)}$            & 4564 & 33 & 863 & 1.39 $\times 10^6$ &24567.518 & 60
[258] & 0.349 $\times 10^6$\\
$C_{2,q,C_F C_A^2 \zeta_3}^{(3)}$    &  284 &  8 &  63 &    26.83 &29.918 & 7 [17]& 30.46 \\
 $C_{2,q,C_F C_A^2 \zeta_4}^{(3)}$    &   48 &  3 &  11 &     0.32 &0.643 & 1 [1]& 1.01\\
 $C_{2,q,C_F C_A^2 \zeta_5}^{(3)}$    &   19 &  2 &   5 &     0.08 &0.175 & 1 [1]& 0.42\\
 $C_{2,q,C_F^2 N_F }^{(3)}$           & 1762 & 20 & 348 & 40237.45 &2339.516 & 28 [107]&
7548.56\\
 $C_{2,q,C_F^2 N_F \zeta_3}^{(3)}$    &   87 &  4 &  21 &     1.94 &2.354 & 3 [5]& 2.83\\
 $C_{2,q,C_F^2 N_F \zeta_4}^{(3)}$    &   15 &  2 &   3 &     0.07 &0.101 & 1 [1]&  0.34\\
 $C_{2,q,C_F C_A N_F }^{(3)}$         & 1847 & 20 & 360 & 47661.64  &2507.362 & 28 [111]& 7525.89\\
 $C_{2,q,C_F C_A N_F \zeta_3}^{(3)}$  &   89 &  4 &  24 &     2.47 &2.935 & 3 [8]& 3.19\\
 $C_{2,q,C_F C_A N_F \zeta_4}^{(3)}$  &   15 &  2 &   3 &     0.06 &0.101 & 1 [1]& 0.34\\
 $C_{2,q,C_F N_F^2}^{(3)}$            &  131 &  5 &  30 &    58.00 &5.347 & 7 [22]& 12.22\\
 $C_{2,q,C_F N_F^2 \zeta_3}^{(3)}$    &   15 &  2 &   3 &     0.06 &0.101 & 1 [1]& 0.38\\
\hline
 $C_{2,q,dabc}^{(3)}$                 & 1199 & 14 & 242 &  6583.27 &738.498 & 15 [62]& 841.24\\
 $C_{2,q,dabc \zeta_3}^{(3)}$         &  109 &  4 &  25 &     2.33 &3.164 & 2[7]& 2.40\\
 $C_{2,q,dabc \zeta_5}^{(3)}$         &    8 &  1 &   2 &     0.03 &0.041 & 0[0]& 0.10\\
\hline \hline
\end{tabular}
\renewcommand{\arraystretch}{1.0}
\end{center}
}
\end{table}
%%%%%%%%%%%%%%%%%%%%%%%%%%%%%%%%%%%%%%%%%%%%%%%%%%%%%%%%%%%%%%%%%%%%%%%%%%%%%
%%%%%%%%%%%%%%%%%%%%%%%%%%%%%%%%%%%%%%%%%%%%%%%%%%%%%%%%%%%%%%%%%%%%%%%%%%%%%
%%%%%%%%%%%%%%%%%%%%  TABLE 6 %%%%%%%%%%%%%%%%%%%%%%%%%%%%%%%%%%%%%%%%%%%%%%%
\begin{table}[ht]
\caption{\sf Run parameters for the unfolding of the unpolarized quarkonic
         Wilson Coefficients for the structure function $F_L(x,Q^2)$.}
%\label{table:TAB:NS}
{\footnotesize
\begin{center}
\renewcommand{\arraystretch}{1.3}
\begin{tabular}{||l|r|r|r|r|r|r|r||}
\hline \hline
\multicolumn{1}{||c|}{ }        &
\multicolumn{1}{c|}{number of}  &
\multicolumn{1}{c|}{order of}   &
\multicolumn{1}{c|}{degree of}  &
\multicolumn{1}{c|}{total time} &
\multicolumn{1}{c|}{length of}  &
\multicolumn{1}{c|}{number of}  &
\multicolumn{1}{c||}{solution}
\\
\multicolumn{1}{||c|}{ }          &
\multicolumn{1}{c|}{terms} &
\multicolumn{1}{c|}{recurrence}   &
\multicolumn{1}{c|}{recurrence}   &
\multicolumn{1}{c|}{[sec]}        &
\multicolumn{1}{c|}{recurrence}   &
\multicolumn{1}{c|}{harm. sums}   &
\multicolumn{1}{c||}{time [sec]}   \\
\multicolumn{1}{||c|}{ } &
\multicolumn{1}{c|}{needed} &
\multicolumn{1}{c|}{ }   &
\multicolumn{1}{c|}{ }   &
\multicolumn{1}{c|}{ }       &
\multicolumn{1}{c|}{[kbyte]}   &
\multicolumn{1}{c|}{a [b] }   &
\multicolumn{1}{c||}{ }   \\
\hline \hline
 $C_{L,q,C_F}^{(1)}$                &   5&  1&   1&0.02  &0.033 &0[0] &0.12 \\
\hline
 $C_{L,q,C_F^2}^{(2)}$              & 203&  5&  51&7.86  &12.381 &6[11] &17.21
\\
 $C_{L,q,C_F^2 \zeta_3}^{(2)}$      &   5&  1&   1&0.02  &0.033 &0[0] & 0/13\\
 $C_{L,q,C_A C_F}^{(2)}$            & 159&  4&  43&4.32  &7.624 &5[8] &
11.43\\
 $C_{L,q,C_A C_F \zeta_3}^{(2)}$    &   5&  1&   1&0.02  &0.033 &0[0] & 0.12\\
 $C_{L,q,C_F N_F}^{(2)}$            &  19&  2&   4&0.05  &0.134 &1[6] & 5.30\\
\hline
$C_{L,q,C_F^3}^{(3)}$               &   2419& 22& 472&110504.41 &4555.679 &27 [142] & 19060.10\\
$C_{L,q,C_F^3 \zeta_3}^{(3)}$       &    131&  5&  34&     3.52 &6.257 &3 [10] & 7.40\\
$C_{L,q,C_F^3 \zeta_5}^{(3)}$       &     11&  1&   3&     0.05 &0.069 &0 [0] & 0.14\\
$C_{L,q,C_F^2 C_A}^{(3)}$           &   2551& 23& 486&124064.39 &5176.054 &27 [144] & 24614.00 \\
$C_{L,q,C_F^2 C_A \zeta_3}^{(3)}$   &    131&  5&  34&     4.51 &6.807 &3 [10] & 7.39\\
$C_{L,q,C_F^2 C_A \zeta_5}^{(3)}$   &     11&  1&   3&     0.05 &0.069 & 0 [0] & 0.14\\
$C_{L,q,C_F C_A^2}^{(3)}$           &   1803& 18& 344& 42500.82 &2064.227 & 27 [109]&6269.33 \\
$C_{L,q,C_F C_A^2 \zeta_3}^{(3)}$   &    131&  5&  31&     3.50 &5.463 & 2 [10]& 6.32\\
$C_{L,q,C_F C_A^2 \zeta_5}^{(3)}$   &     11&  1&   3&     0.05 &0.069 & 0 [0]& 0.15\\
$C_{L,q,C_F C_A N_F}^{(3)}$         &   1014& 14& 203&  4041.82 &539.901 & 13 [58] & 896.70 \\
$C_{L,q,C_F C_A N_F \zeta_3}^{(3)}$ &     41&  2&  12&     0.19 &0.518 & 1 [5]&  0.92\\
$C_{L,q,C_F^2 N_F}^{(3)}$           &    959& 13& 188&  3507.92 &400.784 & 13 [51] & 769.90\\
$C_{L,q,C_F^2 N_F \zeta_3}^{(3)}$   &     29&  2&   8&     0.15 &6.257 & 1 [1]& 0.85\\
$C_{L,q,C_F N_F^2}^{(3)}$           &     47&  3&  10&     1.58 &0.498& 2 [4]& 1.45\\
\hline
$C_{L,q,dabc N_F}^{(3)}$            &    989& 12& 184&  3536.04 &371.269 & 15[60] & 384.00\\
$C_{L,q,dabc N_F \zeta_3}^{(3)}$    &     89&  4&  18&     1.90 &2.034 & 2 [7] & 2.68\\
$C_{L,q,dabc N_F \zeta_5}^{(3)}$    &      5&  1&   1&     0.02 &0.033 & 0 [0]& 0.12\\
\hline \hline
\end{tabular}
\renewcommand{\arraystretch}{1.0}
\end{center}
}
\end{table}
%%%%%%%%%%%%%%%%%%%%%%%%%%%%%%%%%%%%%%%%%%%%%%%%%%%%%%%%%%%%%%%%%%%%%%%%%%%%%
%%%%%%%%%%%%%%%%%%%%%%%%%%%%%%%%%%%%%%%%%%%%%%%%%%%%%%%%%%%%%%%%%%%%%%%%%%%%%
%%%%%%%%%%%%%%%%%%%%  TABLE 7 %%%%%%%%%%%%%%%%%%%%%%%%%%%%%%%%%%%%%%%%%%%%%%%
\begin{table}[ht]
\caption{\sf Run parameters for the unfolding of the unpolarized gluonic
         Wilson Coefficients for the structure function $F_2(x,Q^2)$.}
%\label{table:TAB:NS}
{\footnotesize
\begin{center}
\renewcommand{\arraystretch}{1.3}
\begin{tabular}{||l|r|r|r|r|r|r|r||}
\hline \hline
\multicolumn{1}{||c|}{ }        &
\multicolumn{1}{c|}{number of}  &
\multicolumn{1}{c|}{order of}   &
\multicolumn{1}{c|}{degree of}  &
\multicolumn{1}{c|}{total time} &
\multicolumn{1}{c|}{length of}  &
\multicolumn{1}{c|}{number of}  &
\multicolumn{1}{c||}{solution}
\\
\multicolumn{1}{||c|}{ }          &
\multicolumn{1}{c|}{terms} &
\multicolumn{1}{c|}{recurrence}   &
\multicolumn{1}{c|}{recurrence}   &
\multicolumn{1}{c|}{[sec]}        &
\multicolumn{1}{c|}{recurrence}   &
\multicolumn{1}{c|}{harm. sums}   &
\multicolumn{1}{c||}{time [sec]}   \\
\multicolumn{1}{||c|}{ } &
\multicolumn{1}{c|}{needed} &
\multicolumn{1}{c|}{ }   &
\multicolumn{1}{c|}{ }   &
\multicolumn{1}{c|}{ }       &
\multicolumn{1}{c|}{[kbyte]}   &
\multicolumn{1}{c|}{a [b] }   &
\multicolumn{1}{c||}{ }   \\
\hline \hline
 $C_{2,g, N_F}^{(1)}$                &   24 &  2 &   6 &     0.14 &0.191 &1[3]
&0.72 \\
\hline
 $C_{2,g, N_F C_A}^{(2)}$            &  459 &  9 &  93 &   202.96
&73.022 &7[35] & 70.77 \\
 $C_{2,g, N_F C_A \zeta_3}^{(2)}$    &   8 &  1 &   2 &      0.11 &0.038 &0[0]
& 0.14\\
 $C_{2,g, N_F C_F}^{(2)}$            &  419 &  8 &  91 &   207.98 &63.468 &7[32] & 59.78 \\
 $C_{2,g, N_F C_F \zeta_3}^{(2)}$    &    8 &  1 &   2 &     0.14 &0.038 &0[0]
& 0.14\\

\hline
 $C_{2,g,C_F^2 N_F}^{(3)}$           & 3464 & 28 & 658 &542132.00 &11742.788 &29 [228] &65721.40 \\
 $C_{2,g,C_F^2 N_F \zeta_3}^{(3)}$   &  181 &  6 &  42 &    25.30 &12.171 &3 [14]  &7.67 \\
 $C_{2,g,C_F^2 N_F \zeta_4}^{(3)}$   &   17 &  1 &   5 &     0.23 &0.093 &0 [0] & 0.12\\
 $C_{2,g,C_F^2 N_F \zeta_5}^{(3)}$   &   11 &  1 &   3 &     0.20 &0.067 &0 [0] & 0.15\\
 $C_{2,g,C_A^2 N_F}^{(3)}$           & 4014 & 30 & 739 &869580.00 &16320.095 &28 [261] & 97289.30\\
 $C_{2,g,C_A^2 N_F \zeta_3}^{(3)}$   &  194 &  6 &  44 &    42.39 &13.263 &3 [15] & 8.01\\
 $C_{2,g,C_A^2 N_F \zeta_4}^{(3)}$   &   39 &  2 &  11 &     0.74 &0.408 &1 [3] & 0.67\\
 $C_{2,g,C_A^2 N_F \zeta_5}^{(3)}$   &   11 &  1 &   3 &     0.17 &0.063 &0 [0] & 0.13\\
 $C_{2,g,C_F C_A N_F}^{(3)}$         & 4014 & 30 & 747 &889246.00 &16640.997 &29 [264] & 100830.00\\
 $C_{2,g,C_F C_A N_F \zeta_3}^{(3)}$ &  194 &  6 &  43 &    41.81 &12.999 &3 [15] & 7.90\\
 $C_{2,g,C_F C_A N_F \zeta_4}^{(3)}$ &   39 &  2 &  11 &     0.61 &0.409 &1 [3] & 0.66 \\
 $C_{2,g,C_F C_A N_F \zeta_5}^{(3)}$ &   11 &  1 &   3 &     0.17 &0.068 &0 [0] & 0.11\\
 $C_{2,g,C_F N_F^2}^{(3)}$           & 1553 & 16 & 285 & 22235.00 &1181.805 & 13 [101] &1506.79\\
 $C_{2,g,C_F N_F^2 \zeta_3}^{(3)}$   &   55 &  2 &  16 &     2.81 &0.962 & 1 [3]& 0.70 \\
 $C_{2,g,C_A N_F^2}^{(3)}$           & 1329 & 17 & 259 & 10692.80 &1033.138 &13 [96] & 1162.99\\
 $C_{2,g,C_A N_F^2 \zeta_3}^{(3)}$   &   39 &  2 &  11 &     2.48 &0.666 &1 [3] & 0.70\\
\hline
 $C_{2,g,dabc N_F}^{(3)}$            & 1403 & 15 & 282 & 13951.90 &1048.336 &19 [81] & 2668.66\\
 $C_{2,g,dabc N_f \zeta_3}^{(3)}$    &  142 &  5 &  37 &     8.54 &7.177 & 2 [12] & 6.74\\
 $C_{2,g,dabc N_F \zeta_5}^{(3)}$    &   19 &  1 &   7 &     0.30 &0.139 &0 [0] & 0.14\\
\hline \hline
\end{tabular}
\renewcommand{\arraystretch}{1.0}
\end{center}
}
\end{table}
%%%%%%%%%%%%%%%%%%%%%%%%%%%%%%%%%%%%%%%%%%%%%%%%%%%%%%%%%%%%%%%%%%%%%%%%%%%%%
%%%%%%%%%%%%%%%%%%%%%%%%%%%%%%%%%%%%%%%%%%%%%%%%%%%%%%%%%%%%%%%%%%%%%%%%%%%%%
%%%%%%%%%%%%%%%%%%%%  TABLE 8 %%%%%%%%%%%%%%%%%%%%%%%%%%%%%%%%%%%%%%%%%%%%%%%
\begin{table}[ht]
\caption{\sf Run parameters for the unfolding of the unpolarized gluonic
         Wilson Coefficients for the structure function $F_L(x,Q^2)$.}
%\label{table:TAB:NS}
{\footnotesize
\begin{center}
\renewcommand{\arraystretch}{1.3}
\begin{tabular}{||l|r|r|r|r|r|r|r||}
\hline \hline
\multicolumn{1}{||c|}{ }        &
\multicolumn{1}{c|}{number of}  &
\multicolumn{1}{c|}{order of}   &
\multicolumn{1}{c|}{degree of}  &
\multicolumn{1}{c|}{total time} &
\multicolumn{1}{c|}{length of}  &
\multicolumn{1}{c|}{number of}  &
\multicolumn{1}{c||}{solution}
\\
\multicolumn{1}{||c|}{ }          &
\multicolumn{1}{c|}{terms} &
\multicolumn{1}{c|}{recurrence}   &
\multicolumn{1}{c|}{recurrence}   &
\multicolumn{1}{c|}{[sec]}        &
\multicolumn{1}{c|}{recurrence}   &
\multicolumn{1}{c|}{harm. sums}   &
\multicolumn{1}{c||}{time [sec]}   \\
\multicolumn{1}{||c|}{ } &
\multicolumn{1}{c|}{needed} &
\multicolumn{1}{c|}{ }   &
\multicolumn{1}{c|}{ }   &
\multicolumn{1}{c|}{ }       &
\multicolumn{1}{c|}{[kbyte]}   &
\multicolumn{1}{c|}{a [b] }   &
\multicolumn{1}{c||}{ }   \\
\hline \hline
 $C_{L,g}^{(1)}$                     &  5&  1&  1&0.02  &0.033 &0[0] & 0.13\\
\hline
 $C_{L,g,C_F N_F}^{(2)}$             &153&  4& 38&4.15  &5.941 &2[6] & 5.30\\
 $C_{L,g,C_A N_F}^{(2)}$             &109&  4& 25&1.31  &2.731 &3[10] &4.22 \\
\hline
 $C_{L,g,C_F^2 N_F}^{(3)}$           & 1679 &  17 & 314 & 48496.50 &1498.918 &16 [100] &2019.46 \\
 $C_{L,g,C_F^2 N_F \zeta_3}^{(3)}$   &  120 &   4 &  28 &     3.64 &3.967 &2 [8] & 3.23\\
 $C_{L,g,C_F^2 N_F \zeta_5}^{(3)}$   &    5 &   1 &   1 &     0.02 &0.033 &0 [0] & 0.09\\
%16 [112] & 1632.43\\
 $C_{L,g,C_A^2 N_F}^{(3)}$           & 1671 &  17 & 302 & 29219.30 &1392.205 &16 [112] & 2012.38\\
 $C_{L,g,C_A^2 N_F \zeta_3}^{(3)}$   &  109 &   4 &  24 &     2.46 &3.007 &2[8] &2.836 \\
 $C_{L,g,C_A^2 N_F \zeta_5}^{(3)}$   &    5 &   1 &   1 &     0.03 &0.033 &0 [0] & 0.11\\
 $C_{L,g,C_F C_A N_F}^{(3)}$         & 1935 &  18 & 351 & 44671.90 &2036.550
&16 [116] & 3510.31\\
 $C_{L,g,C_F C_A N_F \zeta_3}^{(3)}$ &  120 &   4 &  28 &     4.43 &4.154 & 2 [8] & 3.10\\
 $C_{L,g,C_F C_A N_F \zeta_5}^{(3)}$ &    5 &   1 &   1 &     0.03 &0.033 & 0 [0]& 0.11\\
 $C_{L,g,C_F N_F^2}^{(3)}$           &  699 &   9 & 140 &  1350.09 &140.949 & 6 [35]& 108.69\\
 $C_{L,g,C_F N_F^2 \zeta_3}^{(3)}$   &   15 &   1 &   4 &     0.06 &0.074 & 0 [0]& 0.17\\
 $C_{L,g,C_A N_F^2 }^{(3)}$          &  419 &   8 &  90 &    526.25 &57.569 & 6 [30]& 47.40\\
 $C_{L,g,C_A N_F^2 \zeta_3}^{(3)}$   &    5 &   1 &   1 &      0.02 &0.033 & 0 [0]& 0.08\\
\hline
 $C_{L,g,dabc N_F}^{(3)}$           & 1109 & 13 & 231 & 10155.40 &618.402 &18 [75] & 1714.70\\
 $C_{L,g,dabc N_F \zeta_3}^{(3)}$   &  129 &  5 &  27 &     2.18 &3.858 &2 [11] &4.09 \\
 $C_{L,g,dabc N_F \zeta_5}^{(3)}$   &   11 &  2 &   2 &     0.06 &0.074 &0 [0] & 0.12\\
\hline \hline
\end{tabular}\renewcommand{\arraystretch}{1.0}
\end{center}
}
\end{table}
%%%%%%%%%%%%%%%%%%%%%%%%%%%%%%%%%%%%%%%%%%%%%%%%%%%%%%%%%%%%%%%%%%%%%%%%%%%%%
%\afterpage{\clearpage} % force tables now.
%\newpage

%%%%%%%%%%%%%%%%%%%%%%%%%%%%%%%%%%%%%%%%%%%%%%%%%%%%%%%%%%%%%%%%%%%%%
\newpage
%%%%%%%%%%%%%%%%%%%%%%%%%%%%%%%%%%%%%%%%%%%%%%%%%%%%%%%%%%%%%%%%%%%%%%%
\section{Appendix: The non-singlet anomalous dimensions}
\label{sec:A}
%%%%%%%%%%%%%%%%%%%%%%%%%%%%%%%%%%%%%%%%%%%%%%%%%%%%%%%%%%%%%%%%%%%%%%%%

\vspace{1mm}
\noindent
The non-singlet anomalous dimensions $P_{qq}^{(0)}$ 
and $P_{qq}^{k,\pm}(n)|_{k=1,2}$ are given by

\begin{eqnarray}
%------------------------------------------------------------------------------
\label{NS1}
P_{qq}^{0}(n) &=& C_F \left[4 S_1-\frac{3 n^2+3 n+2}{n (n+1)}\right] \\
%------------------------------------------------------------------------------
P_{qq}^{1,-}(n) &=& C_F^2 \Biggl[-\frac{3 n^6+9 n^5+9 n^4-5 n^3-24 n^2-32
                          n-24}{2 n^3 (n+1)^3}-16 S_{-3}
\nonumber\\ &&
+S_{-2} \left(\frac{16}{n (n+1)}
                          -32 S_1\right)+S_1 \left(\frac{8 (2 n+1)}{n^2 (n+1)^2}-16 S_2
                          \right)+\frac{4 \left(3 n^2+3 n+2\right)}{n (n+1)} S_2
\nonumber\\ &&
-16 S_3+32 S_{-2,1}+\frac{16 (-1)^n}{(n+1)^3} \Biggr]
\nonumber\\ &+& C_A C_F \Biggl[-\frac{51 n^5+102 n^4+655 n^3+484 n^2+12 n+144}{18 n^3 (n+1)^2}
                          +8 S_{-3}+\frac{268}{9} S_1
\nonumber\\ &&
+S_{-2} \left(16 S_1-\frac{8}{n
(n+1)}
\right)-\frac{44}{3} S_2+8 S_3-16 S_{-2,1}-\frac{8 (-1)^n}{(n+1)^3}\Biggr]
\nonumber\\ &+& C_F N_F \Biggl[\frac{3 n^4+6 n^3+47 n^2+20 n-12}{9 n^2 (n+1)^2}
-\frac{40}{9} S_1+\frac{8}{3} S_2 \Biggr]
\\
%------------------------------------------------------------------------------
P_{qq}^{1,+}(n) &=& C_F^2 \Biggl[-\frac{3 n^6+9 n^5+9 n^4+59 n^3+40 n^2+32 n+8}{2 n^3 (n+1)^3}
-16 S_{-3}
\nonumber\\ &&
+S_{-2} \left(\frac{16}{n (n+1)}-32 S_1\right)
%\nonumber\\ & &
+S_1 \left(\frac{8 (2 n+1)}
{n^2 (n+1)^2}-16 S_2\right)
\nonumber\\ &&
+\frac{4 \left(3 n^2+3 n+2\right)}{n (n+1)}S_2-16 S_3+32
S_{-2,1}+\frac{16 (-1)^n}{(n+1)^3} \Biggr]
\nonumber\\ &+&
C_A C_F \Biggl[-\frac{51 n^5+153 n^4+757 n^3+851 n^2+208 n-132}{18 n^2 (n+1)^3}+8 S_{-3}
+\frac{268}{9} S_1
\nonumber\\ &&
+S_{-2} \left(16 S_1-\frac{8}{n (n+1)}\right)-\frac{44}{3} S_2
+8 S_3-16 S_{-2,1}-\frac{8 (-1)^n}{(n+1)^3} \Biggr]\nonumber\\ &+&
C_F N_F \Biggl[ \frac{3 n^4+6 n^3+47 n^2+20 n-12}{9 n^2 (n+1)^2}-\frac{40}{9} S_1
+\frac{8}{3} S_2 \Biggr]
\\
%------------------------------------------------------------------------------
P_{qq}^{2,-}(n) &=& C_F^3 \Biggl\{\left(\frac{64}{n (n+1)}-128 S_1\right)
S_{-2}^2
+\Biggl(\frac{16 \left(3 n^6+9 n^5+9 n^4+17 n^3+6 n^2+8 n+2\right)}
{n^3 (n+1)^3}
\nonumber\\ &+&
S_1
\left(\frac{64 \left(3 n^2-n+1\right)}{n^2 (n+1)^2}-1408
S_2\right)-\frac{64 \left(3 n^2+3 n
-11\right) S_2}{n (n+1)}+1536 S_3+128 S_{-2,1}
\nonumber\\ &-&
2304 S_{2,1}\Biggr)
S_{-2}-\frac{16
\left(3 n^2+3 n+2\right) S_2^2}{n (n+1)}
-\frac{P_1(n)}
{2 n^5 (n+1)^5}-576 S_{-5}
\nonumber
%\\
\nonumber
\end{eqnarray}
\begin{eqnarray}
&+&
S_{-4} \left(-\frac{16 \left(9 n^2+9 n-26\right)}{n
(n+1)}-832 S_1\right)
\\
&+&
S_{-3} \left(640 S_1^2-\frac{32 \left(3 n^2+3
n+20\right) S_1}{n (n+1)}+\frac{16 \left(21 n^2+17 n+20\right)}{n^2
(n+1)^2}-320 S_{-2}-2240 S_2\right)
\nonumber\\ &+&
(-1)^n \Biggl(-\frac{48 \left(2
n^2-n+1\right)}{(n+1)^5}+\frac{128 S_{-2}}{(n+1)^3}+\frac{96 (5 n+3)
S_1}{(n+1)^4}
-\frac{64 S_2}{(n+1)^3}\Biggr)
\nonumber\\ &+&
\frac{4 \left(13 n^4+26
n^3+13 n^2-16 n-20\right) S_3}{n^2 (n+1)^2}-\frac{16 \left(15 n^2+15
n+2\right) S_4}{n (n+1)}-192 S_5-832 S_{-4,1}
\nonumber\\ &+&
\frac{896 S_{-3,1}}{n
(n+1)}+1152 S_{-3,2}+S_1^2 \left(-\frac{32 \left(3 n^2+3 n+1\right)}{n^3
(n+1)^3}-768 S_{-2,1}\right)-\frac{32 \left(15 n^2+11 n+16\right)
S_{-2,1}}{n^2 (n+1)^2}
\nonumber\\ &+&
S_2 \left(\frac{2 \left(3 n^6+9 n^5+9 n^4+19
n^3+12 n^2-4 n-16\right)}{n^3 (n+1)^3}+64 S_3+2176
S_{-2,1}\right)
\nonumber\\ &+&
\frac{32 \left(3 n^2+3 n-26\right) S_{2,-2}}{n (n+1)}-1472
S_{3,-2}+\frac{64 \left(3 n^2+3 n-2\right) S_{3,1}}{n (n+1)}+192
S_{3,2}+192 S_{4,1}
\nonumber\\ &+&
2304 S_{-3,1,1}+512 S_{-2,1,-2}+\frac{384
\left(n^2+n-4\right) S_{-2,1,1}}{n (n+1)}+S_1 \Biggl(64 S_2^2-\frac{64
(2 n+1) S_2}{n^2 (n+1)^2}
\nonumber\\ &+&
\frac{4 \left(22 n^6+186 n^5+167 n^4-40 n^3-115
n^2-120 n-44\right)}{n^4 (n+1)^4}-192 S_3+64 S_4-1792
S_{-3,1}
\nonumber\\ &-&
\frac{192 \left(n^2+n-4\right) S_{-2,1}}{n (n+1)}+1664
S_{2,-2}+256 S_{3,1}+3072 S_{-2,1,1}\Biggr)+2304 S_{-2,2,1}+2304
S_{2,1,-2}
\nonumber\\ &-&
384 S_{3,1,1}-4608 S_{-2,1,1,1}
\nonumber\\ &+&
\left(C_F^3 - \frac{3}{2} C_F^2 C_A\right) \zeta_3 \Biggl[ -\frac{24 \left(5 n^4+10 n^3+9
n^2+4 n+4\right)}{n^2 (n+1)^2}-192 S_{-2}\Biggr]\Biggr\}
\nonumber\\ &+&
C_A C_F^2 \Biggl\{\left(256 S_1-\frac{16
\left(3 n^2+3 n+8\right)}{n (n+1)}\right) S_{-2}^2
\nonumber\\ &+&
\Biggl[-\frac{8 \left(81
n^6+243 n^5-229 n^4-389 n^3-130 n^2+228 n+72\right)}{9 n^3 (n+1)^3}+\frac{32
\left(31 n^2+31 n-81\right) S_2}{3 n (n+1)}
\nonumber\\ &+&
S_1 \left(1728
S_2-\frac{32 \left(134 n^4+268 n^3+215 n^2+45 n+54\right)}{9 n^2
(n+1)^2}\right)-1792 S_3-192 S_{-2,1}+2688 S_{2,1}\Biggr]
S_{-2}
\nonumber\\ &+&
\frac{176}{3} S_2^2-\frac{P_2(n)}{36 n^5
(n+1)^5}+672 S_{-5}+S_{-4} \left(\frac{8 \left(97 n^2+97 n-210\right)}{3 n
(n+1)}+1120 S_1\right)
\nonumber\\ &+&
S_{-3} \Biggl(-576 S_1^2+\frac{16 \left(31
n^2+31 n+108\right) S_1}{3 n (n+1)}-\frac{8 \left(268 n^4+536 n^3+811
n^2+507 n+450\right)}{9 n^2 (n+1)^2}
\nonumber\\ &+&
480 S_{-2}+2656 S_2\Biggr)+(-1)^n
\Biggl(\frac{8 \left(382 n^2+41 n-161\right)}{9 (n+1)^5}-\frac{256
S_{-2}}{(n+1)^3}-\frac{16 (127 n+121) S_1}{3 (n+1)^4}
\nonumber\\ &+&
\frac{32
S_2}{(n+1)^3}\Biggr)-\frac{8 \left(385 n^4+770 n^3+427 n^2+6 n-126\right)
S_3}{9 n^2 (n+1)^2}+\frac{8 \left(151 n^2+151 n-30\right) S_4}{3 n
(n+1)}
\nonumber\\ &+&
384 S_5+864 S_{-4,1}-\frac{960 S_{-3,1}}{n (n+1)}-1344
S_{-3,2}\nonumber
\end{eqnarray}
\begin{eqnarray}
&+& S_2 \left(\frac{2 \left(453 n^5+906 n^4+1325 n^3+488 n^2-120
n+144\right)}{9 n^3 (n+1)^2}-32 S_3-2624 S_{-2,1}\right)
\nonumber\\ &+&
\frac{16
\left(268 n^4+536 n^3+625 n^2+321 n+414\right) S_{-2,1}}{9 n^2
(n+1)^2}+S_1^2 (128 S_3+896 S_{-2,1})
\nonumber\\ &-&
\frac{16 \left(31 n^2+31
n-174\right) S_{2,-2}}{3 n (n+1)}+1824 S_{3,-2}-\frac{32 \left(29 n^2+29
n-24\right) S_{3,1}}{3 n (n+1)}-384 S_{3,2}-384 S_{4,1}
\nonumber\\ &-&
2688
S_{-3,1,1}-768 S_{-2,1,-2}+S_1 \Biggl(-\frac{8 \left(135 n^6+731 n^5+245
n^4-617 n^3-395 n^2-309 n-144\right)}{9 n^4 (n+1)^4}
\nonumber\\ &-&
\frac{2144}{9}
S_2+\frac{32 \left(31 n^2+31 n-12\right) S_3}{3 n (n+1)}+160
S_4+1920 S_{-3,1}+\frac{32 \left(31 n^2+31 n-84\right) S_{-2,1}}{3 n
(n+1)}
\nonumber\\ &-&
1856 S_{2,-2}-512 S_{3,1}-3584 S_{-2,1,1}\Biggr)-\frac{64 \left(31
n^2+31 n-84\right) S_{-2,1,1}}{3 n (n+1)}-2688 S_{-2,2,1}-2688
S_{2,1,-2}
\nonumber\\ &+&
768 S_{3,1,1}+5376 S_{-2,1,1,1} \Biggr\}
\nonumber\\ &+&
C_A^2 C_F \Biggl[ \left(\frac{24
\left(n^2+n+2\right)}{n (n+1)}-96 S_1\right) S_{-2}^2
+
\Biggl(\frac{8
\left(27 n^6+81 n^5-155 n^4-271 n^3-92 n^2+78 n+27\right)}{9 n^3
(n+1)^3}
\nonumber\\ &+&
S_1 \left(\frac{16 \left(134 n^4+268 n^3+188 n^2+54
n+45\right)}{9 n^2 (n+1)^2}-512 S_2\right)-\frac{32 \left(11 n^2+11
n-24\right) S_2}{3 n (n+1)}+512 S_3
\nonumber\\ &+&
64 S_{-2,1}-768 S_{2,1}\Biggr)
S_{-2}+\frac{P_3(n)}{108 n^5 (n+1)^5}-192
S_{-5}+S_{-4} \left(-\frac{8 \left(35 n^2+35 n-66\right)}{3 n (n+1)}-352
S_1\right)
\nonumber\\ &+&
(-1)^n \left(-\frac{16 \left(82 n^2+17 n-47\right)}{9
(n+1)^5}+\frac{96 S_{-2}}{(n+1)^3}+\frac{16 (41 n+47) S_1}{3
(n+1)^4}\right)
\nonumber\\ &+&
S_{-3} \Biggl(128 S_1^2-\frac{16 \left(11 n^2+11
n+24\right) S_1}{3 n (n+1)}+\frac{8 \left(134 n^4+268 n^3+311 n^2+177
n+135\right)}{9 n^2 (n+1)^2}
\nonumber\\ &-&
160 S_{-2}-768 S_2\Biggr)+\frac{4 \left(389
n^4+778 n^3+398 n^2+9 n-81\right) S_3}{9 n^2 (n+1)^2}-\frac{8 \left(55
n^2+55 n-24\right) S_4}{3 n (n+1)}
\nonumber\\ &-&
160 S_5-224 S_{-4,1}+\frac{256
S_{-3,1}}{n (n+1)}+384 S_{-3,2}+S_1^2 (-64 S_3-256
S_{-2,1})
\nonumber\\ &-&
\frac{16 \left(134 n^4+268 n^3+245 n^2+111 n+135\right)
S_{-2,1}}{9 n^2 (n+1)^2}+S_2 \left(768
S_{-2,1}-\frac{4172}{27}\right)
\nonumber\\ &+&
\frac{16 \left(11 n^2+11 n-48\right)
S_{2,-2}}{3 n (n+1)}-544 S_{3,-2}+\frac{32 \left(11 n^2+11 n-12\right)
S_{3,1}}{3 n (n+1)}
\nonumber\\ &+&
192 S_{3,2}+192 S_{4,1}+768 S_{-3,1,1}+256
S_{-2,1,-2}+\frac{64 \left(11 n^2+11 n-24\right) S_{-2,1,1}}{3 n
(n+1)}
\nonumber\\ &+&
S_1 \Biggl(\frac{2 \left(245 n^8+980 n^7+1542 n^6+1524 n^5+851
n^4+100 n^3+36 n^2+22 n-6\right)}{3 n^4 (n+1)^4}
\nonumber\\ &-&
\frac{8 \left(11 n^2+11
n-8\right) S_3}{n (n+1)}
-128 S_4-512 S_{-3,1}-\frac{32 \left(11
n^2+11 n-24\right) S_{-2,1}}{3 n (n+1)}
\nonumber\\ &+&
512 S_{2,-2}+256 S_{3,1}+1024
S_{-2,1,1}\Biggr)+768 S_{-2,2,1}
+768 S_{2,1,-2,}-384 S_{3,1,1} \nonumber
\end{eqnarray} \begin{eqnarray}
&-& 1536
S_{-2,1,1,1} \Biggr]
\nonumber\\  &+&
C_A^2 C_F \zeta_3 \Biggl[-\frac{12 \left(5
n^4+10 n^3+9 n^2-4 n-4\right)}{n^2 (n+1)^2}-96 S_{-2} \Biggr]
\nonumber\\ &+&
C_F N_F^2  \Biggl[ \frac{51 n^6+153 n^5+57 n^4+35
n^3+96 n^2+16 n-24}{27 n^3 (n+1)^3}-\frac{16}{27} S_1-\frac{80}{27}
S_2+\frac{16}{9} S_3\Biggr]
\nonumber\\
&+& C_F^2 N_F \Bigg[-\frac{32}{3} S_2^2
-\frac{4
\left(15 n^4+30 n^3+79 n^2+16 n-24\right) S_2}{9 n^2 (n+1)^2}
\nonumber\\ &+&
\frac{207
n^8+828 n^7+1443 n^6+1123 n^5-38 n^4-779 n^3-632 n^2+120}{9 n^4
(n+1)^4}-\frac{128}{3} S_{-4}
\nonumber\\ &+&
S_{-3} \left(\frac{32 \left(10 n^2+10
n+3\right)}{9 n (n+1)}-\frac{64}{3} S_1\right)+(-1)^n \left(\frac{64
S_1}{3 (n+1)^3}-\frac{128 (4 n+1)}{9 (n+1)^4}\right)
\nonumber\\ &+&
S_{-2}
\left(-\frac{32 \left(16 n^2+10 n-3\right)}{9 n^2 (n+1)^2}+\frac{640}{9}
S_1-\frac{128}{3} S_2\right)+\frac{16 \left(29 n^2+29 n+12\right)
S_3}{9 n (n+1)}-\frac{128}{3} S_4
\nonumber\\ &+&
S_1 \left(-\frac{2 \left(165
n^5+330 n^4+165 n^3+160 n^2-16 n-96\right)}{9 n^3 (n+1)^2}+\frac{320}{9}
S_2-\frac{128}{3} S_3-\frac{128}{3} S_{-2,1}\right)
\nonumber\\ &-&
\frac{64 \left(10
n^2+10 n-3\right) S_{-2,1}}{9 n (n+1)}+\frac{64}{3} S_{2,-2}+\frac{64}{3}
S_{3,1}+\frac{256}{3} S_{-2,1,1} \Biggr]
\nonumber\\
&+& \left(C_F^2 - C_F C_A \right) N_F \zeta_3 \Biggl[32 S_1-\frac{8
\left(3 n^2+3 n+2\right)}{n (n+1)} \Biggr]
\nonumber\\ &+& C_A C_F N_F \Biggl[
-\frac{2 \left(270 \
n^7+810 n^6-463 n^5-1392 n^4-211 n^3-206 n^2-156 n+144\right)}{27 n^4
(n+1)^3}
\nonumber\\ &+&
\frac{64}{3} S_{-4}+S_{-3} \left(\frac{32}{3} S_1-\frac{16
\left(10 n^2+10 n+3\right)}{9 n (n+1)}\right)+(-1)^n \left(\frac{64 (4
n+1)}{9 (n+1)^4}-\frac{32 S_1}{3 (n+1)^3}\right)
\nonumber\\ &+&
\frac{1336}{27}
S_2+S_{-2} \left(\frac{16 \left(16 n^2+10 n-3\right)}{9 n^2
(n+1)^2}-\frac{320}{9} S_1+\frac{64}{3} S_2\right)-\frac{8 \left(14
n^2+14 n+3\right) S_3}{3 n (n+1)}+\frac{80}{3} S_4
\nonumber\\ &+&
\frac{32 \left(10
n^2+10 n-3\right) S_{-2,1}}{9 n (n+1)}+S_1 \Biggl(-\frac{4 \left(209
n^6+627 n^5+627 n^4+281 n^3+36 n^2+36 n+18\right)}{27 n^3 (n+1)^3}
\nonumber\\ &+&
16
S_3+\frac{64}{3} S_{-2,1}\Biggr)-\frac{32}{3} S_{2,-2}-\frac{64}{3}
S_{3,1}-\frac{128}{3} S_{-2,1,1} \Biggr]
\nonumber\\
%&+& C_A C_F N_F \zeta_3 \Biggl[
%\frac{8 \left(3 n^2+3 n+2\right)}{n (n+1)}-32 S_1
%\Biggr]
\\
P_{qq}^{2,+} &=&
C_F^3 \Biggl[ \left(\frac{64}{n (n+1)}-128 S_1\right)
S_{-2}^2+\Biggl(\frac{16 \left(3 n^6+9 n^5+9 n^4+n^3+2 n^2+4 n+2\right)}{n^3
(n+1)^3} \nonumber\\
&+&
S_1 \left(-\frac{64 \left(3 n^2+7 n+5\right)}{n^2 (n+1)^2}-1408
S_2\right)-\frac{64 \left(3 n^2+3 n-11\right) S_2}{n (n+1)}+1536
S_3+128 S_{-2,1}
\nonumber
\end{eqnarray}
\begin{eqnarray}
&-&
2304 S_{2,1}\Biggr) S_{-2}-\frac{16 \left(3 n^2+3
n+2\right) S_2^2}{n (n+1)}-\frac{P_4(n)}{2 n^5 (n+1)^5}-576
S_{-5}
\nonumber\\ &+&
S_{-4} \left(-\frac{16 \left(9 n^2+9 n-26\right)}{n (n+1)}-832
S_1\right)+S_{-3} \Biggl(640 S_1^2-\frac{32 \left(3 n^2+3 n+20\right)
S_1}{n (n+1)}
\nonumber\\ &+&
\frac{16 \left(9 n^2+5 n+8\right)}{n^2 (n+1)^2}-320
S_{-2}-2240 S_2\Biggr)+(-1)^n \Biggl(\frac{16 \left(2 n^2+11
n+1\right)}{(n+1)^5}+\frac{128 S_{-2}}{(n+1)^3}
\nonumber\\ &+&
\frac{96 (5 n+3)
S_1}{(n+1)^4}-\frac{64 S_2}{(n+1)^3}\Biggr)+\frac{4 \left(13 n^4+26
n^3+13 n^2-16 n-20\right) S_3}{n^2 (n+1)^2}
\nonumber\\ &-&
\frac{16 \left(15 n^2+15
n+2\right) S_4}{n (n+1)}-192 S_5-832 S_{-4,1}+\frac{896 S_{-3,1}}{n
(n+1)}+1152 S_{-3,2}
\nonumber\\ &+&
S_1^2 \left(-\frac{32 \left(3 n^2+3 n+1\right)}{n^3
(n+1)^3}-768 S_{-2,1}\right)-\frac{32 \left(3 n^2-n+4\right) S_{-2,1}}{n^2
(n+1)^2}
\nonumber\\ &+&
S_2 \left(\frac{2 \left(3 n^6+9 n^5+9 n^4+83 n^3+76 n^2+60
n+16\right)}{n^3 (n+1)^3}+64 S_3+2176 S_{-2,1}\right)
\nonumber\\ &+&
\frac{32 \left(3
n^2+3 n-26\right) S_{2,-2}}{n (n+1)}-1472 S_{3,-2}+\frac{64 \left(3 n^2+3
n-2\right) S_{3,1}}{n (n+1)}+192 S_{3,2}+192 S_{4,1}
\nonumber\\ &+&
2304 S_{-3,1,1}+512
S_{-2,1,-2}+\frac{384 \left(n^2+n-4\right) S_{-2,1,1}}{n (n+1)}+S_1
\Biggl(64 S_2^2-\frac{64 (2 n+1) S_2}{n^2 (n+1)^2}
\nonumber\\ &+&
\frac{4 \left(22
n^6-54 n^5+23 n^4+88 n^3+197 n^2+160 n+52\right)}{n^4 (n+1)^4}-192 S_3+64
S_4-1792 S_{-3,1} \nonumber\\ &-& \frac{192 \left(n^2+n-4\right) S_{-2,1}}{n (n+1)}+1664 S_{2,-2}+256 S_{3,1}+3072 S_{-2,1,1}\Biggr)+2304
S_{-2,2,1}
\nonumber\\ &+&
2304 S_{2,1,-2}
-384 S_{3,1,1}-4608 S_{-2,1,1,1} \Biggr]
\nonumber\\ &+&
C_F^3 \zeta_3 \Biggl[ -\frac{24 \left(5 n^4+10
n^3+n^2-4 n-4\right)}{n^2 (n+1)^2}-192 S_{-2} \Biggr]
\nonumber\\
&+& C_A C_F^2 \Biggl\{\left(256 S_1-\frac{16
\left(3 n^2+3 n+8\right)}{n (n+1)}\right) S_{-2}^2
\nonumber\\ &+&
\Biggl(-\frac{8 \left(81
n^5+243 n^4-337 n^3-1181 n^2-526 n-60\right)}{9 n^2 (n+1)^3}+\frac{32
\left(31 n^2+31 n-81\right) S_2}{3 n (n+1)}
\nonumber\\ &+&
S_1 \left(1728
S_2-\frac{32 \left(134 n^4+268 n^3+89 n^2-81 n-72\right)}{9 n^2
(n+1)^2}\right)-1792 S_3-192 S_{-2,1}+2688 S_{2,1}\Biggr)
S_{-2}
\nonumber\\ &+&
\frac{176}{3} S_2^2-\frac{P_5(n)}{36 n^4 (n+1)^4}+672
S_{-5}+S_{-4} \left(\frac{8 \left(97 n^2+97 n-210\right)}{3 n (n+1)}+1120
S_1\right)
\nonumber\\ &+&
S_{-3} \Biggl(-576 S_1^2+\frac{16 \left(31 n^2+31
n+108\right) S_1}{3 n (n+1)}-\frac{8 \left(268 n^4+536 n^3+487 n^2+183
n+126\right)}{9 n^2 (n+1)^2}
\nonumber
\end{eqnarray}\begin{eqnarray}
&+&
480 S_{-2}+2656 S_2\Biggr)+(-1)^n
\left(\frac{8 (346 n-125)}{9 (n+1)^4}-\frac{256 S_{-2}}{(n+1)^3}-\frac{16
(103 n+73) S_1}{3 (n+1)^4}+\frac{32 S_2}{(n+1)^3}\right)
\nonumber\\ &-&
\frac{8
\left(385 n^4+770 n^3+427 n^2+6 n-126\right) S_3}{9 n^2 (n+1)^2}+\frac{8
\left(151 n^2+151 n-30\right) S_4}{3 n (n+1)}+384 S_5
\nonumber\\ &+&
864
S_{-4,1}-\frac{960 S_{-3,1}}{n (n+1)}-1344 S_{-3,2}+S_2 \Biggl(\frac{2
\left(453 n^5+1359 n^4+2231 n^3+1525 n^2+80 n-264\right)}{9 n^2 (n+1)^3}
\nonumber\\ &-&
32
S_3-2624 S_{-2,1}\Biggr)+\frac{16 \left(268 n^4+536 n^3+301 n^2-3
n+90\right) S_{-2,1}}{9 n^2 (n+1)^2}+S_1^2 (128 S_3+896
S_{-2,1})
\nonumber\\ &-&
\frac{16 \left(31 n^2+31 n-174\right) S_{2,-2}}{3 n (n+1)}+1824
S_{3,-2}-\frac{32 \left(29 n^2+29 n-24\right) S_{3,1}}{3 n (n+1)}-384
S_{3,2}-384 S_{4,1}
\nonumber\\ &-&
2688 S_{-3,1,1}-768 S_{-2,1,-2}+S_1
\Biggl(-\frac{8 \left(135 n^6-649 n^5-1039 n^4-569 n^3+487 n^2+621
n+216\right)}{9 n^4 (n+1)^4}
\nonumber\\ &-&
\frac{2144}{9} S_2+\frac{32 \left(31 n^2+31
n-12\right) S_3}{3 n (n+1)}+160 S_4+1920 S_{-3,1}+\frac{32 \left(31
n^2+31 n-84\right) S_{-2,1}}{3 n (n+1)}
\nonumber\\ &-&
1856 S_{2,-2}-512 S_{3,1}-3584
S_{-2,1,1}\Biggr)-\frac{64 \left(31 n^2+31 n-84\right) S_{-2,1,1,n}}{3 n
(n+1)}-2688 S_{-2,2,1}
\nonumber\\ &-&
2688 S_{2,1,-2}+768 S_{3,1,1}+5376 S_{-2,1,1,1}
\Biggr]
\nonumber\\ &+& C_A C_F^2 \zeta_3 \Biggl[\frac{36 \left(5
n^4+10 n^3+n^2-4 n-4\right)}{n^2 (n+1)^2}+288 S_{-2} \Biggr]
\nonumber\\ &+&
C_A^2 C_F  \left(\frac{24 \left(n^2+n+2\right)}{n (n+1)}-96S_1\right)S_{-2}^2+\Biggl(\frac{8 \left(27 n^6+81 n^5-209 n^4-595 n^3-272
  n^2-48 n-9\right)}{9 n^3 (n+1)^3}\nonumber\\ &+&S_1 \left(\frac{16 \left(134
  n^4+268 n^3+116 n^2-18 n-27\right)}{9 n^2 (n+1)^2}-512 S_2\right)
-\frac{32 \left(11 n^2+11 n-24\right) S_2}{3 n (n+1)}+512 S_3
\nonumber\\ &+&
64 S_{-2,1}-768
S_{2,1}\Biggr) S_{-2}
+\frac{P_6(N)}
{108 n^3 (n+1)^5}-192 S_{-5}+S_{-4} \left(-\frac{8 \left(35 n^2+35
  n-66\right)}{3 n (n+1)}-352 S_1\right)
\nonumber\\ &+&
(-1)^n \left(-\frac{16 \left(91 n^2+80 n-29\right)}{9 (n+1)^5}+\frac{96
  S_{-2}}{(n+1)^3}+\frac{16 (29 n+23) S_1}{3 (n+1)^4}\right)
\nonumber\\ &+&
S_{-3} \Biggl(128 S_1^2-\frac{16 \left(11 n^2+11 n+24\right) S_1}{3 n
  (n+1)}+\frac{8 \left(134 n^4+268 n^3+203 n^2+69 n+27\right)}{9 n^2 (n+1)^2}
\nonumber\\ &-&
160 S_{-2}-768 S_2\Biggr)+\frac{4 \left(389 n^4+778 n^3+398 n^2+9 n-81\right)
  S_3}{9 n^2 (n+1)^2}-\frac{8 \left(55 n^2+55 n-24\right) S_4}{3 n (n+1)}
\nonumber\\ &-&
160 S_5-224 S_{-4,1}+\frac{256 S_{-3,1}}{n (n+1)}+384 S_{-3,2}+S_1^2 (-64
S_3-256 S_{-2,1})
\nonumber\\ &-&
\frac{16 \left(134 n^4+268 n^3+137 n^2+3 n+27\right) S_{-2,1}}{9 n^2
  (n+1)^2}+S_2 \left(768 S_{-2,1}-\frac{4172}{27}\right)
\nonumber\\ &+&
\frac{16 \left(11 n^2+11 n-48\right) S_{2,-2}}{3 n (n+1)}-544
S_{3,-2}+\frac{32 \left(11 n^2+11 n-12\right) S_{3,1}}{3 n (n+1)}+192 S_{3,2}
\nonumber
\end{eqnarray}\begin{eqnarray}
&+&
192 S_{4,1}+768 S_{-3,1,1}+256 S_{-2,1,-2}+\frac{64 \left(11 n^2+11
  n-24\right) S_{-2,1,1}}{3 n (n+1)}
\nonumber\\ &+&
S_1 \Biggl(\frac{2 \left(245 n^8+980 n^7+1542 n^6+964 n^5+211 n^4-60 n^3+156
  n^2+222 n+90\right)}{3 n^4 (n+1)^4}
\nonumber\\ &-&
\frac{8 \left(11 n^2+11 n-8\right) S_3}{n (n+1)}-128 S_4-512 S_{-3,1}
\nonumber\\ &-&
\frac{32 \left(11 n^2+11 n-24\right) S_{-2,1}}{3 n (n+1)}+512
S_{2,-2}+256 S_{3,1}+1024 S_{-2,1,1}\Biggr)+768 S_{-2,2,1}
\nonumber\\ &+&
768
S_{2,1,-2}-384 S_{3,1,1}-1536 S_{-2,1,1,1} \Biggr\}
\nonumber\\ &+& C_A^2 C_F \zeta_3 \Biggl[ -\frac{12 \left(5
n^4+10 n^3+n^2-4 n-4\right)}{n^2 (n+1)^2}-96 S_{-2} \Biggr]
\nonumber\\
&+& C_F^2 N_F  \Biggl\{ -\frac{32}{3} S_2^2-\frac{4
\left(15 n^4+30 n^3+79 n^2+16 n-24\right) S_2}{9 n^2 (n+1)^2}+\frac{
P_7(n)}{9 n^4
(n+1)^4}-\frac{128}{3} S_{-4}
\nonumber\\ &+&
S_{-3} \left(\frac{32 \left(10 n^2+10
n+3\right)}{9 n (n+1)}-\frac{64}{3} S_1\right)+(-1)^n \left(\frac{64
S_1}{3 (n+1)^3}-\frac{128 (4 n+1)}{9 (n+1)^4}\right)
\nonumber\\ &+&
S_{-2}
\left(-\frac{32 \left(16 n^2+10 n-3\right)}{9 n^2 (n+1)^2}+\frac{640}{9}
S_1-\frac{128}{3} S_2\right)+\frac{16 \left(29 n^2+29 n+12\right)
S_3}{9 n (n+1)}-\frac{128}{3} S_4
\nonumber\\ &+&
S_1 \left(-\frac{2 \left(165
n^5+495 n^4+495 n^3+517 n^2+336 n+80\right)}{9 n^2 (n+1)^3}+\frac{320}{9}
S_2-\frac{128}{3} S_3-\frac{128}{3} S_{-2,1}\right)
\nonumber\\ &-&
\frac{64 \left(10
n^2+10 n-3\right) S_{-2,1}}{9 n (n+1)}+\frac{64}{3} S_{2,-2}+\frac{64}{3}
S_{3,1}+\frac{256}{3} S_{-2,1,1} \Biggr\}
\nonumber\\ &+& C_F^2 N_F \zeta_3 \Biggl[32 S_1-\frac{8
\left(3 n^2+3 n+2\right)}{n (n+1)}\Biggr]
\nonumber\\ &+& C_F N_F^2  \Biggl[ \frac{51 n^6+153 n^5+57 n^4+35
n^3+96 n^2+16 n-24}{27 n^3 (n+1)^3}-\frac{16}{27} S_1-\frac{80}{27}
S_2+\frac{16}{9} S_3 \Biggr]
\nonumber\\ &+&
C_A C_F N_F \Biggl[ -\frac{2 \left(270
n^7+1080 n^6+383 n^5-979 n^4-571 n^3+507 n^2+106 n-132\right)}{27 n^3
(n+1)^4}
\nonumber\\ &+&
\frac{64}{3} S_{-4}+S_{-3} \left(\frac{32}{3} S_1-\frac{16
\left(10 n^2+10 n+3\right)}{9 n (n+1)}\right)+(-1)^n \left(\frac{64 (4
n+1)}{9 (n+1)^4}-\frac{32 S_1}{3 (n+1)^3}\right)
\nonumber\\ &+&
\frac{1336}{27}
S_2+S_{-2} \left(\frac{16 \left(16 n^2+10 n-3\right)}{9 n^2
(n+1)^2}-\frac{320}{9} S_1+\frac{64}{3} S_2\right)-\frac{8 \left(14
n^2+14 n+3\right) S_3}{3 n (n+1)}+\frac{80}{3} S_4
\nonumber\\ &+&
\frac{32 \left(10
n^2+10 n-3\right) S_{-2,1}}{9 n (n+1)}+S_1 \Biggl(-\frac{4 \left(209
n^6+627 n^5+627 n^4+137 n^3-108 n^2-108 n-54\right)}{27 n^3 (n+1)^3}
\nonumber\\ &+&
16 S_3+\frac{64}{3} S_{-2,1}\Biggr)-\frac{32}{3} S_{2,-2}-\frac{64}{3}
S{_3,1}-\frac{128}{3} S_{-2,1,1}\Biggr]
\nonumber\\ &+& C_A C_F N_F \zeta_3 \Biggl[\frac{8
\left(3 n^2+3 n+2\right)}{n (n+1)}-32 S_1\Biggr] \nonumber
\end{eqnarray}\begin{eqnarray}
\label{NS2}
P_{qq}^{2,-,dabc} &=& \frac{d_{abc} d^{abc}}{N_c} N_F  \Biggl[
-\frac{P_8(n)}{3 n^5 (n+1)^5 (n+2)^3}+\frac{4
\left(n^2+n+2\right) S_{-3}}{n^2 (n+1)^2}
-\frac{P_9(n) S_1}{3
n^4 (n+1)^4 (n+2)^3}
\nonumber\\ &+&
S_{-2} \left(-\frac{8 S_1
\left(n^2+n+2\right)^2}{(n-1) n^2 (n+1)^2 (n+2)}-\frac{4 \left(n^6+3 n^5-8
n^4-21 n^3-23 n^2-12 n-4\right)}{(n-1) n^3 (n+1)^3 (n+2)}\right)
\nonumber\\ &+&
(-1)^n
\Biggl(\frac{16 \left(5 n^6+29 n^5+78 n^4+118 n^3+114 n^2+72 n+16\right)
S_1}{3 (n-1) n^2 (n+1)^3 (n+2)^3}
\nonumber\\ &-&
\frac{4 \left(13 n^8+74 n^7+179 n^6+314
n^5+644 n^4+1000 n^3+816 n^2+352 n+64\right)}{3 (n-1) n^3 (n+1)^4
(n+2)^3}\Biggr)
\nonumber\\ &-&
\frac{2 \left(n^2+n+2\right) S_3}{n^2 (n+1)^2}-\frac{8
\left(n^2+n+2\right) S_{-2,1}}{n^2 (n+1)^2} \Biggr]
\end{eqnarray}
For brevity we abbreviated $S_{\vec{a}}(n) \equiv S_{\vec{a}}$. Here,
$C_A = N_c, C_F = (N_c^2-1)/(2 N_c)$ are $SU(N_c)$ color factors, $N_F$ denotes the
number of quark flavors
and $N_c$ is the number of colors, with $N_c = 3$ for Quantum Chromodynamics.
We have accounted for the color factor $T_R = 1/2$ explicitly, which is the 
same for
all groups $SU(N_c)$. $d_{abc}$ denotes a $SU(N_c)$ structure constant and the
Einstein convention is applied calculating $d_{abc} d^{abc}$.

\vspace*{1mm}\noindent
The functions $P_i(n)$ which appear in Eqs.~(\ref{NS1}--\ref{NS2}) are
given by
\begin{eqnarray}
%----------------------------------------------------------------------------
P_1(n) &=& 29 n^{10}+145 n^9+130 n^8-146 n^7-479 n^6-11 n^5-464 n^4-1748 n^3
-1600 n^2
\nonumber\\ & &
-752 n-16\\
P_2(n) &=& 1359 n^{10}+6795 n^9+15246 n^8+15646
n^7+3851 n^6-35089 n^5-34648 n^4
\nonumber\\ &&
+12280 n^3+32592 n^2+17616 n+3456
\\
P_3(n) &=&
4971 n^{10}+24855 n^9+11770 n^8-86322 n^7-150929 n^6-135893
n^5-85692 n^4
\nonumber\\ &+&
-18992 n^3+22824 n^2+15840 n+259
\\
P_4(n) &=& 29 n^{10}+145 n^9+226 n^8+110 n^7+353
n^6+501 n^5+976 n^4+940 n^3+576 n^2
\nonumber\\ &&
+208 n+32
\\
P_5(n) &=& 1359 n^8+5436 n^7+8274 n^6+24524 n^5+11103 n^4+12528 n^3+4120 n^2
\nonumber\\ &&
-2560 n-1584
\\
P_6(n) &=&
4971 n^8+24855 n^7+10762 n^6-57138 n^5-92033 n^4-40901 n^3+10692 n^2
\nonumber\\ &&
+1216 n-2904
\\
P_7(n) &=& 207 n^8+828 n^7+1491 n^6+2291 n^5+1338 n^4+453 n^3-8 n^2-160 n-72
\\
P_8(n) &=&
4 \bigl(13 n^{10}+97 n^9+326 n^8+720 n^7+1399 n^6+2416 n^5+3017
n^4+2412 n^3
\nonumber\\ &&
+1184 n^2+336 n+48\bigr)
\\
P_9(n) &=&
2 \bigl(9 n^9+41 n^8-3
n^7-505 n^6-1719 n^5-2951 n^4-3092 n^3-2032 n^2
\nonumber\\ &&
-768 n-144\bigr)
\end{eqnarray}
%----------------------------------------------------------------------------------------

%---------------------------------------------------------------------------------
%---------------------------------------------------------------------------------
\end{document}